\documentclass[sigconf, nonacm]{acmart}

\newcommand\vldbavailabilityurl{https://github.com/HPI-Information-Systems/Mondrian}

\usepackage{graphicx}
\usepackage{balance}  
\usepackage{booktabs} 
\usepackage{color}
\usepackage{soul}
\usepackage{xspace}
\usepackage{url}
\usepackage{float}
\usepackage{enumitem}
\usepackage{algorithm}
\usepackage{ifthen}
\usepackage{algpseudocode}
\algtext*{EndWhile}
\algtext*{EndIf}
\algtext*{EndFor}

\usepackage{efbox}
\efboxsetup{linecolor=black,linewidth=0.5pt}

\usepackage{amsthm}

\usepackage{stackengine}
\usepackage{subcaption}
\usepackage{lipsum}
\usepackage{multirow}
\usepackage[multiple]{footmisc}
\usepackage{xcolor}
\usepackage{booktabs, tabularx}
\usepackage{numprint}
\usepackage{etoolbox}

\begin{document}
\title{Detecting Layout Templates in Complex Multiregion Files}

\author{Gerardo Vitagliano}
\email{gerardo.vitagliano@hpi.de}
\affiliation{%
  \institution{Hasso Plattner Institute, University of Potsdam, Germany}
  \streetaddress{Prof.-Dr.-Helmert-Str. 2-3}
}

\author{Lan Jiang}
\email{lan.jiang@hpi.de}
\affiliation{%
  \institution{Hasso Plattner Institute, University of Potsdam, Germany}
  \streetaddress{Prof.-Dr.-Helmert-Str. 2-3}
}

\author{Felix Naumann}
\email{felix.naumann@hpi.de}
\affiliation{%
  \institution{Hasso Plattner Institute, University of Potsdam, Germany}
  \streetaddress{Prof.-Dr.-Helmert-Str. 2-3}
}

\newcommand{\true}{true}
\newcommand{\false}{false}
\newcommand{\PaperLongVersion}{\true} 

\newcommand{\PaperLong}[1]{%
\ifdefstring{\PaperLongVersion}{\true}{%
{\color{black}#1}}{}}

\newcommand{\PaperShort}[1]{%
\ifdefstring{\PaperLongVersion}{\false}{#1}{}}

\newcommand{\PaperLongFigure}[1]{%
\ifthenelse{\equal{\PaperLongVersion}{\true}}{#1}{}}

\newcommand{\PaperShortFigure}[1]{%
\ifthenelse{\equal{\PaperLongVersion}{\false}}{#1}{}}

\newcommand{\lannote}[1]{\hl{\footnote{\hl{LJ: #1}}}}
\newcommand{\gerardonote}[1]{\hl{\footnote{\hl{GV: #1}}}}
\newcommand{\felixnote}[1]{\hl{\footnote{\hl{FN: #1}}}}

\newcommand{\tabels}[1]{\mathcal{E}}
\newcommand{\tabel}[1]{\e}

\newcommand{\dataset}[1]{\texttt{#1}}
\newcommand{\usecase}[1]{\medskip\noindent\emph{#1}\medskip}

\newcommand{\prep}[1]{\texttt{#1}} 
\newcommand{\term}[1]{\textit{#1}} 

\newtheorem{challenge}{Challenge}

\theoremstyle{definition}
\newtheorem{defn}{Definition}
\newtheorem{task}{Task}

\newcommand{\dpsys}{Data-Knoller}
\newcommand{\multifile}{multiregion\xspace }
\newcommand{\Multifile}{Multiregion\xspace}
\newcommand{\systemname}{Mondrian\xspace }
\newcommand{\projectname}{Mondrian\xspace }

\newcommand{\DECO}{\textsc{Deco}\xspace}
\newcommand{\FUSTE}{\textsc{Fuste}\xspace}

\ifthenelse{\equal{\PaperLongVersion}{\true}}{
\newcommand{\change}[1]{
    {\color{black}#1}%
}}{
\newcommand{\change}[1]{
    {\color{blue}#1}%
}}

\begin{abstract}
Spreadsheets are among the most commonly used file formats for data management, distribution, and analysis.
Their widespread employment makes it easy to gather large collections of data, but their flexible canvas-based structure makes automated analysis difficult without heavy preparation.
One of the common problems that practitioners face is the presence of multiple, independent regions in a single spreadsheet, possibly separated by repeated empty cells.
We define such files as ``\multifile'' files.
In collections of various spreadsheets, we can observe that some share the same layout.
We present the \systemname approach to automatically identify layout templates across multiple files and systematically extract the corresponding regions.
Our approach is composed of three phases: first, each file is rendered as an image and inspected for elements that could form regions; then, using a clustering algorithm, the identified elements are grouped to form regions; finally, every file layout is represented as a graph and compared with others to find layout templates.
We compare our method to state-of-the-art table recognition algorithms on two corpora of real-world enterprise spreadsheets.
Our approach shows the best performances in detecting reliable region boundaries within each file and can correctly identify recurring layouts across files.
\end{abstract}

\maketitle

\ifdefempty{\vldbavailabilityurl}{}{
\vspace{.3cm}
\begingroup\small\noindent\raggedright\textbf{Artifact Availability:}\\
The source code, data, and/or other artifacts have been made available at \url{\vldbavailabilityurl}.
\endgroup
}

\section{Structural File Templates}
\label{sec:intro}

Data comes in all shapes and forms.
The recent blossom of open data portals has made large quantities of spreadsheet files available for public consumption~\cite{openCSV2016, euses2005, fuse2015}.
It is common knowledge that much human time and effort in data-oriented workflows is spent on preparing data files.
Even spreadsheets that are meant for distribution and analysis can be affected by data quality issues and human-induced errors that make information extraction difficult~\cite{enron2015, chiticariuEnterpriseInformationExtraction2010}: they are often used as canvases in which data is spread out in multiple, independent regions with a custom layout and without a well-defined tabular format.
In many cases, there are multiple tables, but metadata regions are also common, e.g., spreadsheet titles, comment sections, or notes to data cells.

\begin{figure}[t]
\centering
\includegraphics[width=0.8\columnwidth]{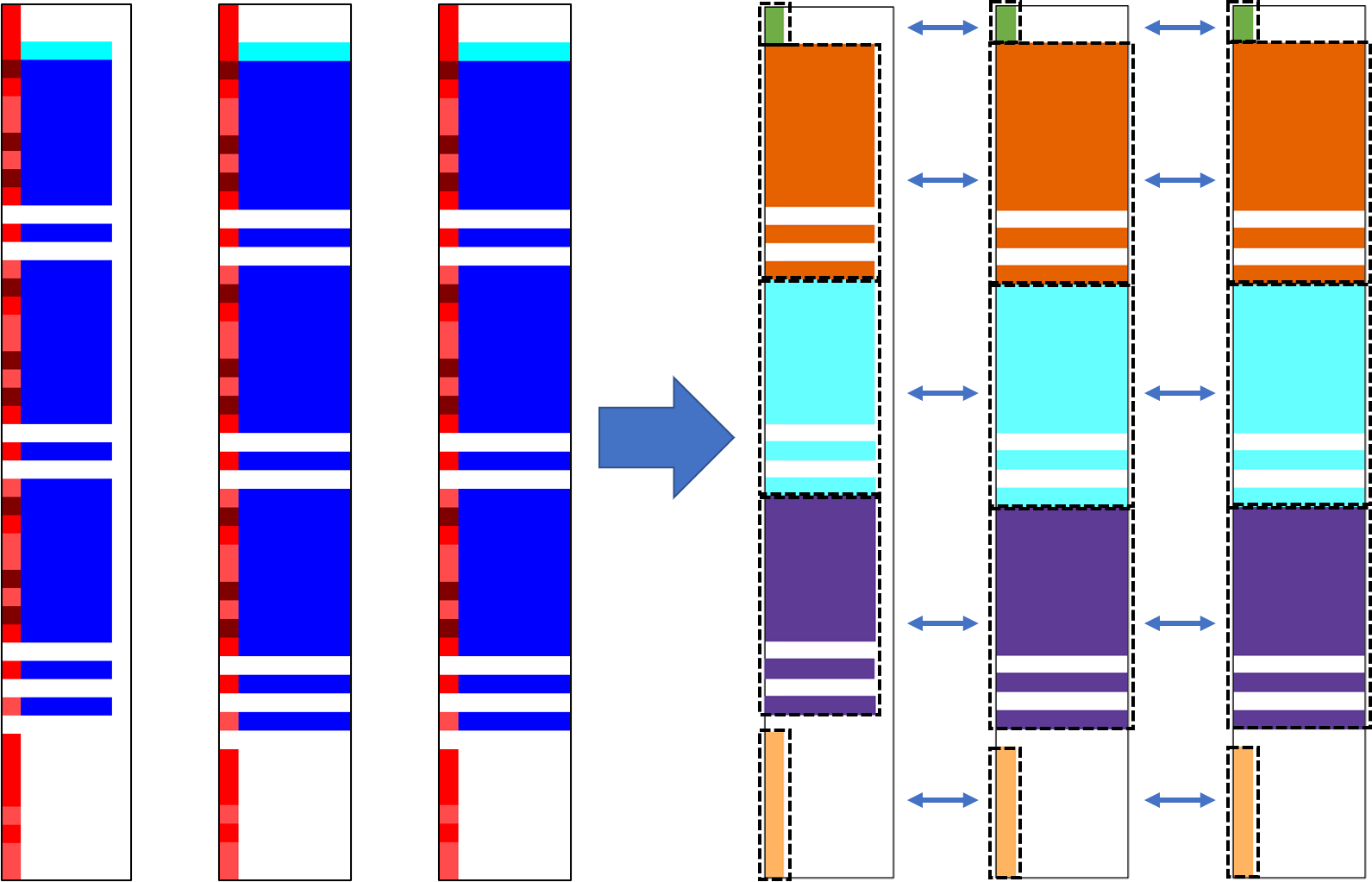}
\caption{
\change{Visual rendering of three different files sharing the same \multifile layout.}
\vspace{-10pt}
}
\label{fig:overview}
\end{figure}
\begin{figure*}[t]
\centering
\includegraphics[width=.8\textwidth]{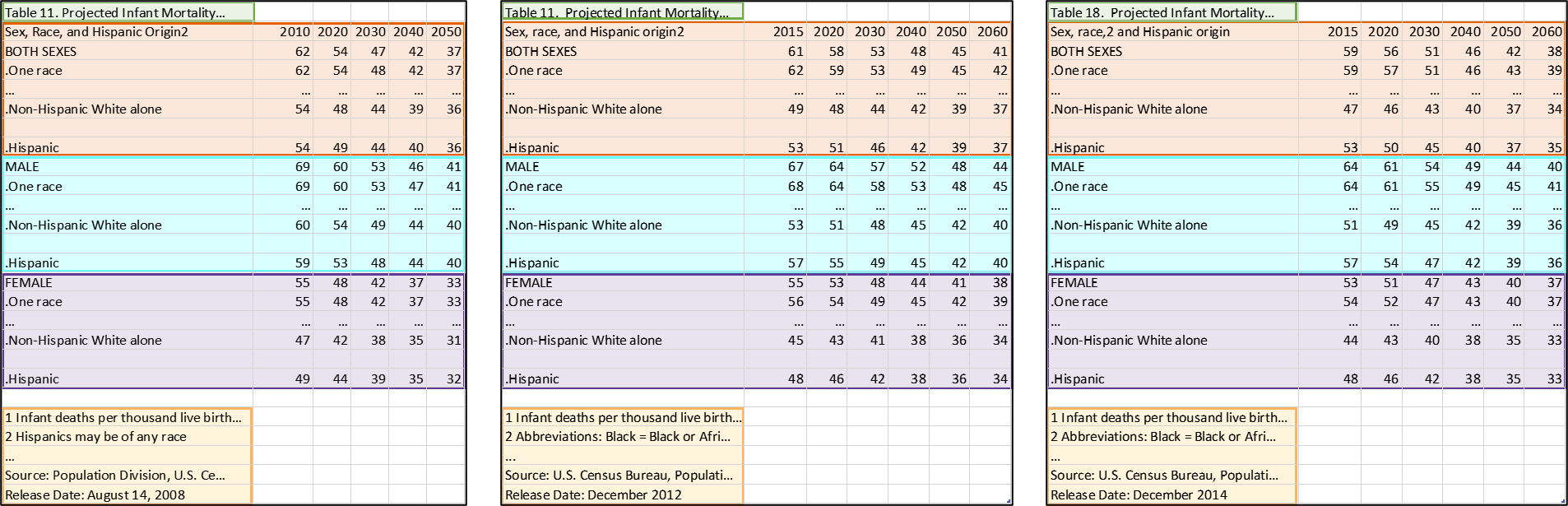}
\caption{Detailed view of the three US Census files \change{of Figure~\ref{fig:overview}} sharing the same \multifile layout.}
\label{fig:overview-detail}
\end{figure*}
As an example, Figure~\ref{fig:overview} depicts the visual structure of three different spreadsheet files from the FUSE corpus~\cite{fuse2015}: they all contain the same three tables (albeit with different data points), a title, and a footnote region, arranged in the same layout.
\change{
A detailed view of their content is shown in Figure~\ref{fig:overview-detail}.}
Due to missing values and empty rows, it is difficult not only to draw the correct table boundaries within one file, but also to recognize that the three files share the same layout.
\change{Such multi-region files are found in enterprise data lakes or open data repositories, often without proper metadata or provenance, and cause these to grow into unordered collections of heterogeneous data~\cite{nargesian2019datalake}.
Often, multiple files are produced in repeatedly automated pipelines, or stem from the same business processes, and therefore share the same ``layout template''.
We define a template to encompass the number, layout, and schema of tables and metadata regions in a file.
Therefore, files from the same template contain related data from the same domain, whether originating from the same source or from multiple connected sources.
In data-oriented workflows, recognizing layout templates serves multiple use cases:
\begin{enumerate}[nosep]
    \item Exploring the content of a data lake, users can discover all files that contain semantically related information to a given input file.
    \item Performing data preparation on a set of files that share a template, users can automatically target the same region in all files irrespective of its file-specific boundaries.
    \item Integrating files from multiple sources, users can use templates as metadata indicating data provenance, and decide to exclude the tables of a template if it is deemed as containing conflicting or poor-quality data.
\end{enumerate}

In Section~\ref{sec:usecase} we present a full-fledged example for the data preparation use case~(2) using the files of Figure~\ref{fig:overview}.

Recent research has addressed the problem of discovering related tabular data in data lakes~\cite{koutrasSIPBFLBK21, zhangI20, zhuDNM19}, but these methods, designed for relational tables, are unfit for \multifile files: first, because the same regions may appear in different positions across different files (e.g., due to slight layout differences); second, because spreadsheets may contain more than just a single table (e.g., preamble or footnotes cells, or multiple tables). 
Therefore, the first step to identify layout templates is to correctly detect and recognize layout regions.
}

Different methods have been proposed for single table spreadsheets~\cite{chenCRF, koci2019genetic, pytheas20}, yet, ``complex'' multiregion layouts are a common occurrence across spreadsheet data sources.
In the \DECO dataset~\cite{koci2019deco}, an annotated sample of 854 files from the ENRON Excel corpus, almost 75\% of the sheets (621) contain more than one region
with 71 layouts recurring in more than one file; in a randomly sampled subset of 886 files from the FUSE spreadsheet dataset, annotated by the authors of this paper, almost half of them (391) show multiple regions, with 31 recurring layouts; and Mitlöhner et al.\ reported that, out of 141k csv files retrieved from open data portals, roughly 3\% of the correctly parsed files contained more than one table, and 4.6\% of those that could not be correctly parsed were showing ``too many tables''~\cite{openCSV2016}.
What is more, previous approaches for automated table extraction in spreadsheets usually rely on format-specific style features.
However, files are more often shared in .csv format.
For example, of 15,497 files distributed on the UK open data portal (\url{data.gov.uk}), 
44.18\% are in .csv format, compared to 
8.81\% in an Excel-specific format (.xls/.xlsx).
The same trend is true for the US open data portal (\url{data.gov}), where out of 192,335 datasets, 9.61\% have a ``csv'' tag, while only 3.19\% have an ``excel'' tag.
Therefore, we design our approach, \systemname, to be general with respect to file format, ignoring rich-text features as encoded in Excel files.
While additional metadata, such as file and/or sheet names could also prove useful for template detection, we observe that these can be unreliable and/or unavailable in real-world scenarios (consider, e.g., how often sheets are labeled ``Sheet1'', or files are machine-named).
Our intuition is to leverage the visual distribution and the literal content of individual cells by converting each file into an image and segment it to find heterogeneous regions: first, we graphically identify individual segments of adjacent data, and then we partition them to have finer-grained elements to cluster together.
Once regions have been detected, file layouts are described as graphs and compared using a similarity flooding-based algorithm to find layout templates.
The graphical rendering of a template inspired us to name our approach after the abstract painter Piet Mondrian.
In proposing \systemname, we make the following contributions:
\begin{enumerate}[nosep]
\item An unsupervised approach that leverages a novel mapping between spreadsheets and the visual image domain to detect and match different regions in spreadsheet files.

\item A framework to analyze and compare \multifile spreadsheets, using a graph representation with an associated similarity algorithm to detect layout templates.

\item A publicly available dataset of structural annotations for 886 spreadsheets, classifying the position and purpose of their composing regions, and a set of template annotations for two datasets, summing up to a total of above 1500 files, identifying classes of files with the same layout.

\item A comprehensive set of experiments to prove the effectiveness of the \systemname approach in solving the region detection and template inference problems, evaluating, and comparing it with state-of-the-art automated methods.


\end{enumerate}
The code artifacts and dataset files are publicly available at the project page \url{\vldbavailabilityurl}.

\section{A Data Preparation Use Case}

\label{sec:usecase}
Consider the historical population data of the United States Census, made publicly available through an open data portal\footnote{\url{https://www2.census.gov/programs-surveys/popproj/tables/} (accessed 25/05/21)}.
The data from each year is summarized in
different tables contained in csv/spreadsheet files, and although some tables are unique to specific years, others recur in multiple years. The files that contain the same tables all share the same layout: they have similar title and footnote cells, and all their tables (when more than one) have the same schema (cf.\ Figure~\ref{fig:overview-detail}).

Consider the three files in Figure~\ref{fig:overview-detail}, containing data about projected infant mortality (some rows excluded for visual clarity). 
All have three tables, a title, and a footnote region, arranged with the same layout.
However, there are slight differences in the files across years.
For example, in the footnote region, the last cell reflects the year, and sometimes cells have different content while the semantic meaning is the same (E.g., ``Source: Population Division, U.S. Census Bureau'' and ``Source: U.S. Census Bureau, Population division'').
The tables themselves have a different number of columns across files, and their headers are updated. 
Finally, the table title also changes from ``Table 11'' to ``Table 18''.
Nonetheless, it is obvious at a glance that the three files come from the same layout template.
With manual human inspection and domain knowledge, it is possible to consolidate tables from the same templates into a single source of truth to enable downstream tasks, 
after some necessary data preparation steps, e.g., remove the footnote lines.
\change{%
Note that within the same template, regions may have slightly different positions in different files: for example footnotes appear from line 41-47 in one of the files and in lines 43-49 in another.
Because of this,
}
without \systemname, preparation steps must be carried out manually for each file, becoming more and more cumbersome and time-consuming the larger the set of input files.
With \systemname, it is possible to leverage the recurring structure of the templates and prepare all template files at once.
In the US Census example, out of 99 spreadsheets, in a few minutes, our system identifies the layout of every file and groups them into fifteen different templates.
For example, for the three files of Figure~\ref{fig:overview-detail}, \systemname detects the region boundaries for each file layout, identifies that all layouts belong to the same template, and determines that the regions across different files are equivalent.
Using the results of \systemname, end-users may perform template-wide transformations, for example
deleting all title and footnote regions, separating the tables, and removing all empty rows 
\change{without having to manually specify the individual region boundaries for all files.}  




\section{Describing \Multifile~Layouts}
\label{sec:multitables}



Before describing the details of our solution, we provide definitions for the concepts of \multifile files, layouts, and layout templates.
Typically, \multifile files can be found in comma-separated values format (\emph{.csv}) or Microsoft Excel format (\emph{.xls/.xlsx}). 
Complex layouts with multiple regions are a byproduct of spreadsheet software rendering data on ``canvases'' where users freely lay out different data (and possibly metadata).\footnote{Some formats and tools allow a spreadsheet to have more than one ``worksheet''. Without loss of generality, we consider each worksheet as a separate file.}
Here, we formalize the concepts needed to describe the layout of \multifile files,
formulate a hierarchy of equivalence notions to compare regions and layouts, and 
state the research problems addressed by our approach.


\subsection{Multiregion spreadsheets}
\label{subsec:definitions}
Our sources of data are spreadsheets, defined as value-delimited files that contain data in cells with a grid structure. 
We assume no specific row- or column-based structure of the content.
We assign each cell a unique identifier $(x,y)$, where $x,y \in \mathbb{N}_0$ correspond to the column and row indices, respectively.
These $(x, y)$ coordinates are points in a Euclidean space with its origin in the top-left corner, in analogy to spreadsheet design.
Every cell serves some purpose in the spreadsheet.
We consider three fundamental types of cells: 
\begin{defn}[Cell types]
A cell $c$ of a spreadsheet $S$ belongs to one of the following mutually disjoint cell types: 
\begin{enumerate}
\item \textbf{Data}, if it carries the data values of a file;
\item \textbf{Metadata}, if its information is related to a set of data cells;
\item \textbf{Empty}, if it does not contain any data or only whitespace characters, e.g., it is used for visual formatting.
\end{enumerate}
\end{defn}
Elements are simple structures, grouping cells of the same type:
\begin{defn}[Element]
\label{def:element}
Given a spreadsheet file $S$, an element $e$ is a rectangular set of adjacent cells of $S$ of the same type.
The element type of $e$ corresponds to the cell type of its cells.
\end{defn}

According to its position in the spreadsheet, an element can be described with the vector $(x_0, y_0, x_1, y_1)$ $\in \{\mathbb{N}_0\}^4$, where the coordinates $(x_0, y_0)$ represent an element's top-left cell and $(x_1, y_1)$ its bottom-right cell. 
Since elements are groups of adjacent cells, in a given spreadsheet we can identify several of them and describe their spatial relationships.
Considering the elements' rectangular nature and the grid-like space of spreadsheets, we encode the relationship between two elements with three features: alignment direction, alignment magnitude, and distance.
The alignment direction is based on the overlap of the elements' projection on the x-axis and the y-axis:
\begin{defn}[Alignment]
Two elements $a:=(a_{x_0},a_{y_0},a_{x_1},a_{y_1})$, $ b:=(b_{x_0},b_{y_0},b_{x_1},b_{y_1})$ are aligned:
    $$\begin{cases}
    \emph{Vertically ($V$) } &\mbox{if } 
    max(a_{y_0}, b_{y_0}) \leq min(a_{y_1},b_{y_1}) 
    \\
    \emph{Horizontally ($H$) } &\mbox{if } 
    max(a_{x_0}, b_{x_0}) \leq min(a_{x_1},b_{x_1})
    \\
    \emph{Not aligned ($N$) } 
    &\mbox{otherwise}
    \end{cases}$$
\end{defn}

It is worthwhile noting that, as they are adjacent groups of cells, the areas of any two given elements in a spreadsheet cannot overlap.
The alignment magnitude is the number of shared points across the axis in the case of horizontal or vertical alignment:
\begin{defn}[Alignment magnitude]
\label{def:alignment-magnitude}
The alignment magnitude between elements $a,b$ is:
    $$\begin{cases}
    \min(a_{y_1},b_{y_1})-\max(a_{y_0}, b_{y_0}) +1
    &\mbox{if } \textit{alignment}(a,b) = V
    \\
    \min(a_{x_1},b_{x_1}) - \max(a_{x_0}, b_{x_0}) +1
    &\mbox{if } \textit{alignment}(a,b) = H
    \\
    0 &\mbox{otherwise}
    \end{cases}$$
\end{defn}

The distance between the elements is calculated as the distance of their two closest points. 
In case the two elements are horizontally or vertically aligned, this resolves to the distance between their closest boundaries; otherwise, it is calculated as the Euclidean distance of the two closest corners:


\begin{defn}[Distance]
\label{def:distance}
The distance $d(a,b)$ between elements is

    $\begin{cases}
    d_v:|\min(a_{x_1},b_{x_1}) - \max(a_{x_0}, b_{x_0})+1|
    & \mbox{if } \textit{alignment}(a,b) = V
    \\
    d_h:|\min(a_{y_1},b_{y_1})  - \max(a_{y_0}, b_{y_0}) + 1|
    & \mbox{if } \textit{alignment}(a,b) = H
    \\
    \sqrt{d_v^2+d_h^2}
    &\mbox{otherwise}
    \end{cases}$
\end{defn}


Often, especially in spreadsheets with complex cell layouts, even non-adjacent cells could be logically grouped.
For example, a table may have missing values that result in empty rows in-between valid data rows (As seen in Figure~\ref{fig:overview}).
Elements are therefore not sufficient to completely describe the layout of a spreadsheet, and we need a higher-order abstraction to group semantically related elements, which are not necessarily adjacent to each other.
Groups of elements can serve different purposes: examples are tables, preambles, footnotes, or any other domain-specific construct.
To abstract their specific purpose, we identify them as regions:
\begin{defn}[Region]
\label{def:region}
A region $R$ is a complete graph having as nodes a set of semantically related, non-empty elements $\mathcal{E}$, connected with edges labeled with their pairwise spatial relationships.
\end{defn}

\begin{figure}[t]
\centering
\includegraphics[width=0.8\linewidth]{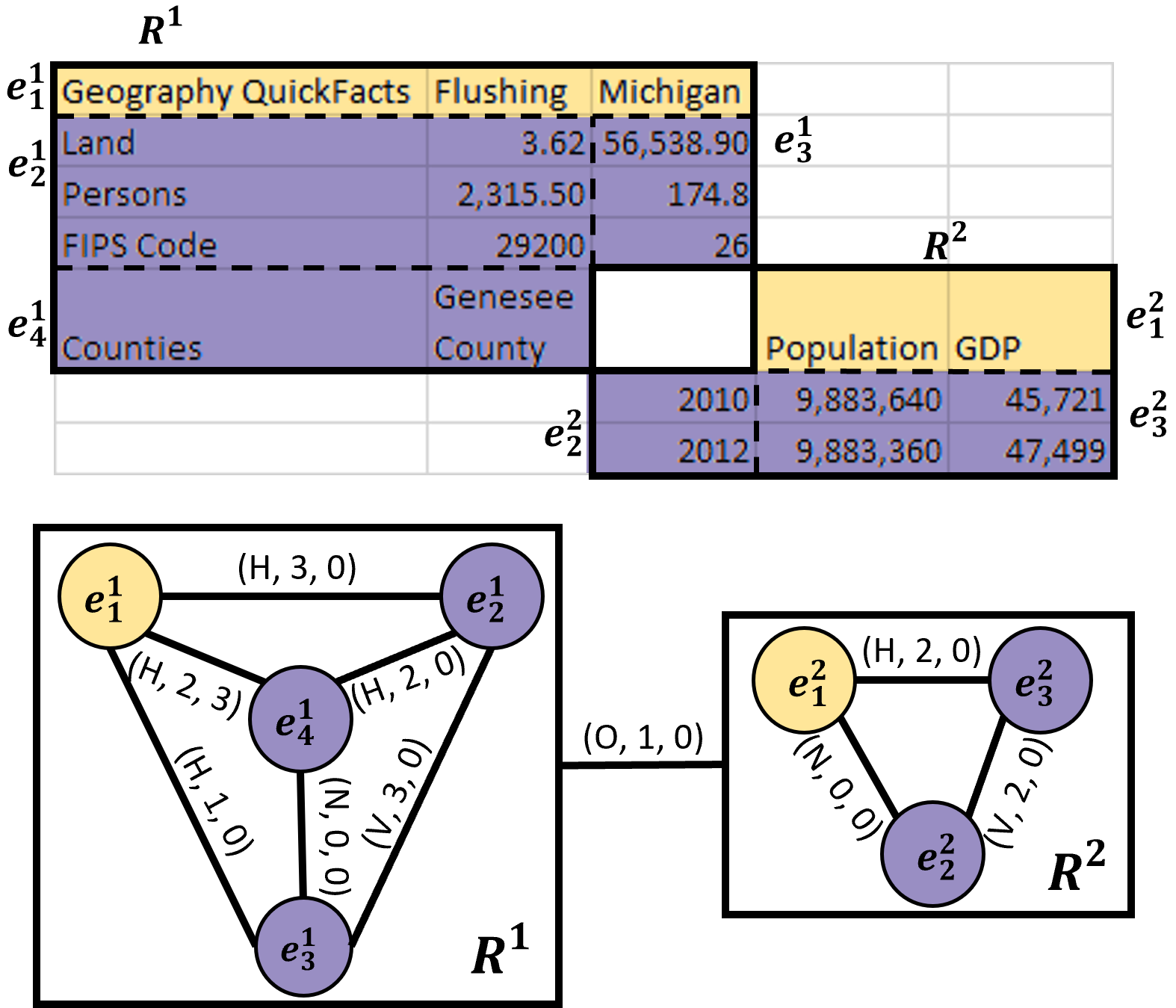}\caption{\change{Two overlapping regions and their graph layout (yellow: metadata elements, purple: data elements).}}
\label{fig:graph}
\end{figure}

\change{
Figure~\ref{fig:graph} shows two given regions, their composing elements and their associated graph layouts: in region $R^1$ the header element $e^1_1$ is horizontally aligned to the two data elements $e^1_2$ and $e^1_4$, and the two data element $e^1_2$ and $e^1_3$ are vertically aligned; in region $R^2$ the data element $e^2_3$ is horizontally aligned to the header element $e^2_1$ and vertically aligned to the data element $e^2_2$.
}



Considering the definition of regions, a \multifile spreadsheet is trivially defined as a spreadsheet containing multiple regions.

Ultimately, our goal is to find structural similarity across different, possibly \multifile files.
To do so, it is first important to identify a ``meaningful'' set of regions for each file: that is to say, draw the boundaries of different regions such that they are independent and serve distinct purposes.
To describe the coordinates of a region boundary in the spreadsheet space, we use the bounding box of its set of elements:
\begin{defn}[Region boundary]
\label{def:bbox}
The boundary of a region $R$, with its elements $\mathcal{E}$, is defined as a rectangle $(x_0, y_0, x_1, y_1)$, where:
$$ x_0 = \min_{ e \in \mathcal{E}} e_{x_0} ,\quad y_0 = \min_{e \in \mathcal{E}} e_{y_0}, \quad x_1 = \max_{e \in \mathcal{E}} e_{x_1},\quad y_1 = \max_{e \in \mathcal{E}} e_{y_1} $$
\end{defn}

Once regions have been identified, we are concerned with their layout.
We extend to pairs of regions the spatial relationship feature vector defined for pairs of elements, using the $(x_0, y_0, x_1, y_1)$ coordinates of region boundaries to compute alignment direction, magnitude, and distance.
One caveat is that considering their boundaries, two given regions can, in general, have overlapping bounding boxes,
which is not the case for elements.
We extend the spatial relationship feature vector for overlapping regions as:
\begin{defn}[Overlapping regions]
Given two regions,
$A:=(a_{x_0},a_{y_0},a_{x_1},a_{y_1})$ and $B:=(b_{x_0},b_{y_0},b_{x_1},b_{y_1})$, their alignment direction is ``overlapping'' (O) if $\max(a_{y_0}, b_{y_0}) \leq \min(a_{y_1},b_{y_1})$ and $\max(a_{x_0}, b_{x_0}) \leq \min(a_{x_1},b_{x_1})$.
Then, the alignment magnitude is
    $(\min(a_{y_1},b_{y_1}) - \max(a_{y_0}, b_{y_0})+1)\cdot(
    \min(a_{x_1},b_{x_1}) - \max(a_{x_0}, b_{x_0})+1)$
 and the distance is~0.
\end{defn}

The magnitude corresponds to the area of the overlap, which ultimately equals the product of the horizontal and vertical alignment magnitudes, considering that two overlapping regions are both horizontally and vertically aligned.
\change{
For example, the two regions $R^1$ and $R^2$ in Figure~\ref{fig:graph} overlap for one cell, and their spatial relationship vector is ('O', 1, 0).
}
Finally, describing a set of non-empty regions with a complete graph, we can define the layout of a spreadsheet:
\begin{defn}[Spreadsheet layout]
\label{def:file-layout}
The layout of a spreadsheet file $S$ is a complete graph having as nodes its set of non-empty regions, connected with edges labeled with their pairwise spatial relationship.
\end{defn}

\subsection{Templates as recurring structures}
\label{subsec:recurring}

Often, region and file layouts are not one-off models, but stem from a systematic creation process.
For example, the US Census open data portal contains the same data report for multiple geographical entities, each downloadable as a separate csv file\footnote{https://www.census.gov/quickfacts/, last accessed May. 31, 2021}.
Our goal is to provide a framework to define and analyze \emph{templates}, i.e., classes of structural equivalence across multiple files.
We compose a hierarchy of equivalence notions, beginning with 
the finest-grained unit of comparison, the cell, and extend it to elements:
\begin{defn}[Cell equivalence]
Two cells $c_1$, $c_2$ are equivalent if their type and content are equal.
Two empty cells are always equivalent.
\end{defn}


\begin{defn}[Element equivalence]
Two elements $e_1$, $e_2$ are equivalent if their types are the same and if there is a one-to-one equivalence between their cells, regardless of their position.
Two empty elements are always equivalent.
\end{defn}

Similar to cells, we consider empty elements equivalent, regardless of their shape, as their purpose is to provide visual information about region layout to end-users.
Recalling Definition~\ref{def:region}, this information is encoded within the attributes of the edges (i.e., the spatial relationship between nodes) of a region graph.
We define element equivalence to be insensitive of cell position to be able to match elements that have equal content differing only in their layout, e.g., two tables with the same column in a different position.
To define region equivalence, we must also be able to include regions with equal structure but different data values, e.g., two tables with the same schema but different data.
\begin{defn}[Region equivalence]
\label{def:region-equivalence}
Two regions $R_1$, $R_2$ are equivalent if there is a one-to-one equivalence between their metadata nodes and their graphs are isomorphic.
\end{defn}

At the spreadsheet level, the definition for layout is similar:
\begin{defn}[Layout equivalence]
Two layouts $L_1$, $L_2$ are equivalent if there is a one-to-one equivalence between their regions and their graphs are isomorphic.
\end{defn}


In practice, if many files are collected from different sources, we want to be able to discover entire sets of equivalent spreadsheets:
\begin{defn}[Layout template]
A layout template $\mathcal{L}$ is a class of equivalent file layouts.
\end{defn}

Recognizing templates is of great value for data preparation, as it potentially saves users the time to manually inspect and prepare individual files: a pipeline of preparation steps can be defined once and executed repeatedly on different files from the same template.
\change{As computing exact graph isomorphism is computationally expensive, \systemname uses approximate similarity metrics to find templates, described in Sections~\ref{subsec:region-detection} and \ref{subsec:layout-similarity}.}

\subsection{Automated layout inference}
\label{subsec:problem}
Given the definitions stated, the problem of recognizing and matching \multifile spreadsheet layouts is composed of several distinct sub-problems that have an inherently visual nature.
The first fundamental problem is to find the correct region boundaries.
A human expert would solve this task by understanding the semantics of the data as well as its spatial distribution.
Then, to identify recurring layouts, they would be required to manually inspect and compare each separate file looking at its data -- a cumbersome, error-prone, and time-consuming task.
According to our definitions of equivalence,
this task requires semantic concepts and possibly domain knowledge, e.g., to distinguish table schemata.
However, to design a general and domain-independent approach, we focus only on structural properties.
\change{We present the \systemname approach to address the following research problem:

\noindent\textbf{Problem Statement:} Given a set of spreadsheet files $\mathcal{F}$, each with its layout $L_f$:
\begin{enumerate}
\item Given a file $f$, how can we determine the set  $R_f$ of regions that compose its layout $L_f$?

\item Given two different regions $r_x$, $r_y$, how can we approximate their equivalence without semantic information?

\item Given pairs of files $f_x$, $f_y \in \mathcal{F}$, how can we measure the similarity of their layouts and use these similarities to recognize unique layout templates $\mathcal{L}$ that occur in $\mathcal{F}$?

\end{enumerate}

}







\section{The Mondrian Approach}
\label{sec:algorithm}

\begin{figure}[b]
\captionsetup[subfigure]{labelformat=simple, labelsep=period}
\captionsetup[subfigure]{width=1.1\linewidth}%
\subcaptionbox{A file of the ENRON corpus viewed in a spreadsheet software. {\label{fig:csv} \vspace{1em}
}}
{
\includegraphics[width=\linewidth]{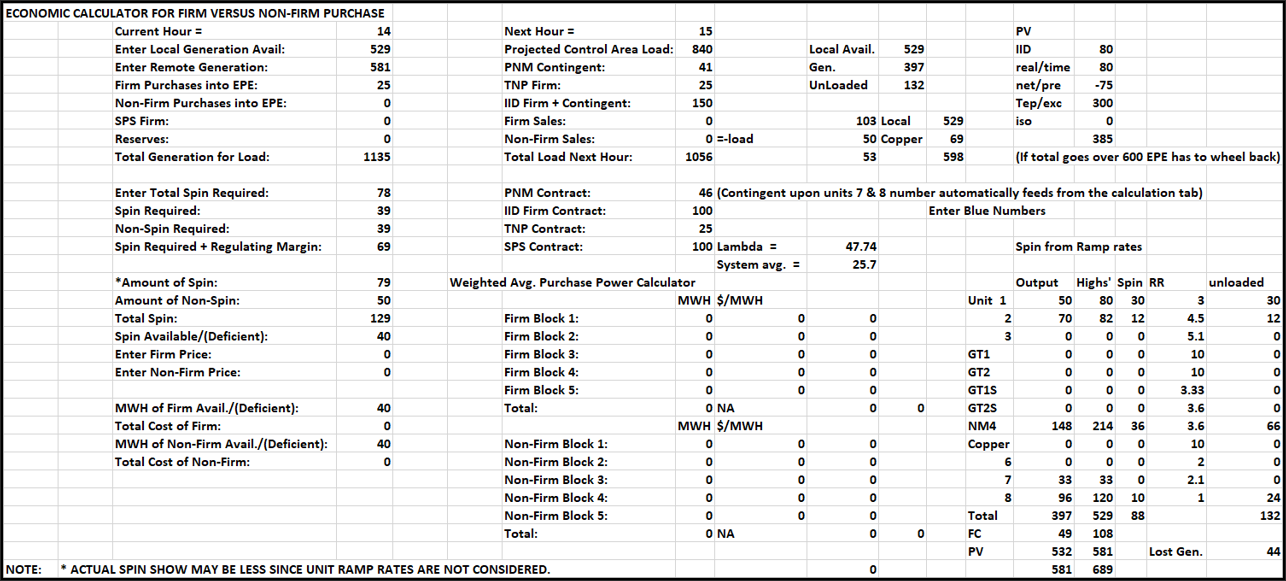}}
\subcaptionbox{Spreadsheet image parsing. \label{fig:rendering-color}}
{
\includegraphics[width=0.4\linewidth]{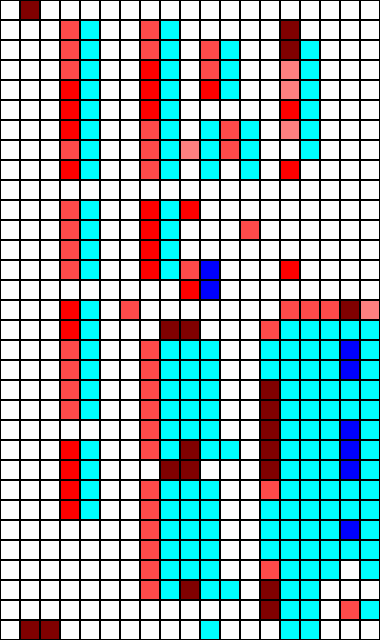}}
\hspace{2em}
\subcaptionbox{Connected components.\label{fig:rendering-binary}}
    {
\includegraphics[width=0.4\linewidth]{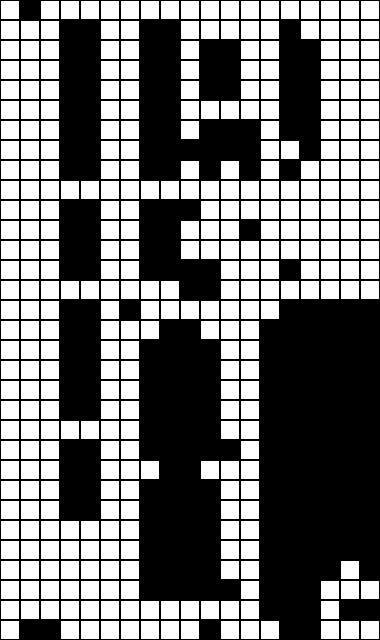}}
\caption{Core intuition of Mondrian -- transposing a spreadsheet to the image domain.}
\label{fig:rendering}
\end{figure}

To identify the conceptual entities defined in Section~\ref{subsec:definitions} in practice, without resorting to semantic knowledge, the intuition of \systemname is to transform the domain of spreadsheets from data content to image.
We convert cells into pixels, encoding their syntactical types into colors.
Then, we find elements by segmenting the file images with a partitioning algorithm and cluster them to detect region boundaries. 
Once regions are identified, we analyze their structural properties and use a similarity measure to match regions across different files.
If two (or more) files are found to have similar regions, we measure the similarity between the graph representations for their layouts and possibly group them into a template.


\subsection{Image parsing and segmentation}
\label{subsec:image-parsing}

To cover the most general cases, our approach takes as input comma-separated value files.
Files with different delimiters or formatted with XML markup, such as Microsoft Excel files, can be easily converted into a `.csv' file.
Ignoring possible markup information is the trade-off for a method applicable to a wide spectrum of spreadsheets, independent of their format specifications.

For native csv files, we cannot assume that all rows have the same number of delimiters.
Thus, we pad rows with empty cells up to the length of the longest row.
Given a csv file with $M$ rows and $N$ columns, we create an image with the dimensions $M \times N$, 
where each pixel represents a cell in the csv file.
Our definitions of entities and their equivalence build upon the concept of ``cell type'': in practice, we substitute semantic types with \textit{syntactic types} and, correspondingly, relax their equivalences into an approximate \emph{structural similarity}.

We identify four fundamental syntactic types: \textit{number}, \textit{datetime}, \textit{string}, \textit{empty}. Except for \textit{empty}, each of these types can be further refined in sub-types: a \textit{number} can be \textit{integer} or \textit{floating-point}; a \textit{datetime} can be a \textit{time} or a \textit{date}; a string can be either \textit{uppercase}, \textit{lowercase}, \textit{titlecase} or \textit{generic}.
In parsing the spreadsheet as an image, we transform every cell into a pixel with a different color according to its type (cf.\ Figures~\ref{fig:csv} and~\ref{fig:rendering-color}).
Table~\ref{tab:datatypes} shows the color corresponding to each data type and a sample cell from Figure~\ref{fig:csv}\footnote{Except for the time and date types, which were not present in the original file.} that was parsed according to that type. 

\begin{table}[h]\centering\small
\begin{tabular}{llll}
Type & Sub-type & Sample cell & Color \\
\toprule
Empty & Empty & `` '' & White \\ 
\midrule
Number & Integer & ``14'' & Light Blue \\
    & Floating-point & ``47.74'' & Dark Blue \\
\midrule
Datetime & Time & ``17:00'' & Light Green \\
    & Date & ``17/9/20'' & Dark Green \\
\midrule
String & Uppercase & ``MWH'' & Maroon \\
    & Lowercase & ``real/time'' & Salmon Red \\
    & Titlecase & ``Firm Sales'' & Tomato Red \\
    & Generic & ``System avg. ='' & Scarlet Red \\
\bottomrule
\end{tabular}
\caption{Data types and their colors.}
\vspace{-20pt}
\label{tab:datatypes}
\end{table}
Recognizing the syntactic type of cells without semantic knowledge is, in general, a coarse-grained and error-prone operation: consider the uncertain nature of the value ``1990'', which can be a date or a number.
As our experiments in Section~\ref{subsec:evaluation-template} demonstrate, however, a coarse-grained parsing is sufficient to approximate region equivalence for the task of template inference, with the reasonable assumption that any parsing mistake would be reflected across all similar files.
To segment the file into elements, we first find connected components, which reflect cell aggregates
that could not be so easily recognized in a spreadsheet software view 
(Figure~\ref{fig:rendering-binary}).
The change in width/height proportion happens because each cell occupies one square pixel in the image, while in the spreadsheet software cell columns and rows can have different widths or heights, usually set according to the length of their values.
With this ``cell normalization'', for example, a human observer is more likely to note the four aligned vertical elements on the left of the image.

\begin{figure}[b]
\captionsetup{justification=centering}
\captionsetup[subfigure]{labelformat=simple, labelsep=period}
\captionsetup[subfigure]{width=1.1\linewidth}%
\centering
\begin{subfigure}{\linewidth}
\includegraphics[width=\linewidth]{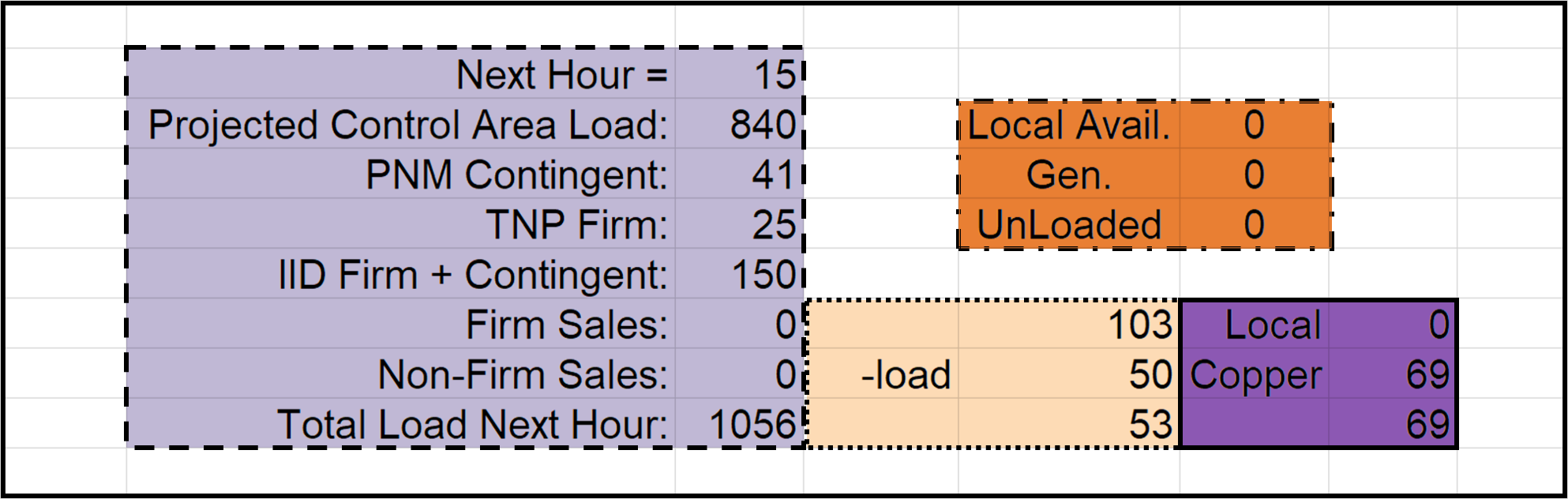}
\caption{Detail of Figure~\ref{fig:csv} highlighting adjacent, independent regions.}
\label{fig:partition-csv}
\end{subfigure}
\par\bigskip
\begin{subfigure}[t]{0.3\linewidth}
\includegraphics[width=\linewidth]{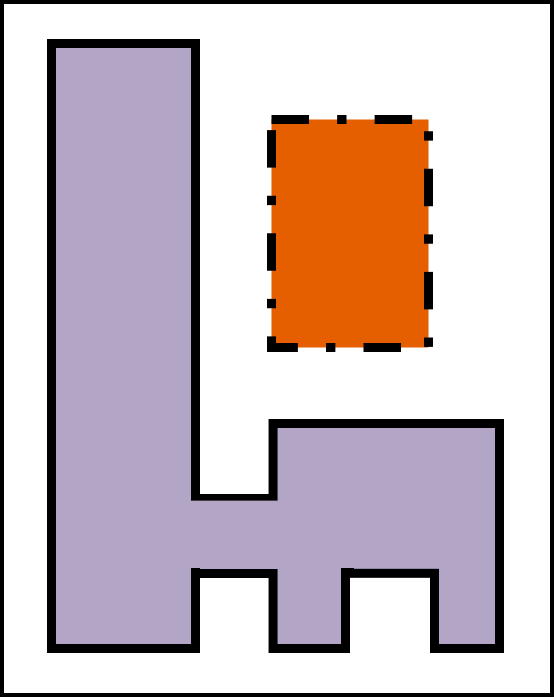}
\caption{Connected \\components.}
\label{fig:partition-coarse}
\end{subfigure}
\hfill{}
\begin{subfigure}[t]{0.3\linewidth}
\includegraphics[width=\linewidth]{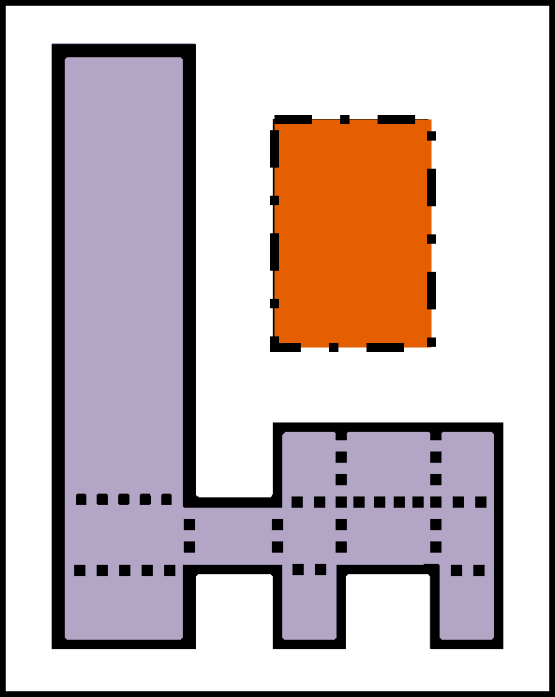}
\caption{Cutting into finer grained elements.}
\label{fig:partition-cut}
\end{subfigure}
\hfill{}
\begin{subfigure}[t]{0.3\linewidth}
\includegraphics[width=\linewidth]{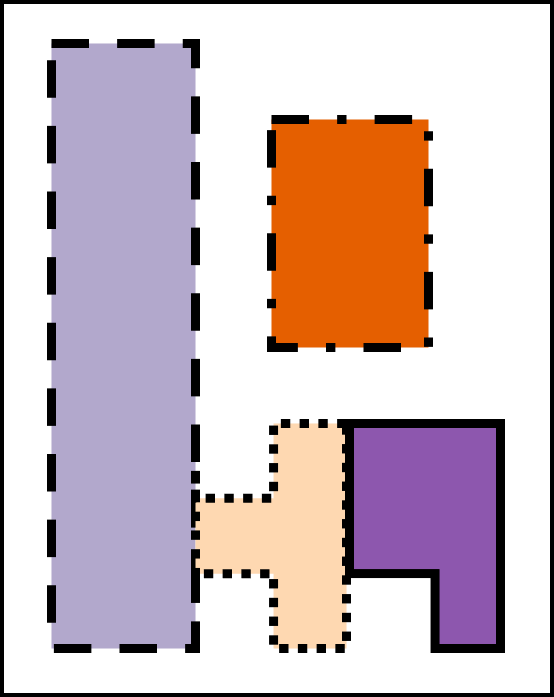}
\caption{Clustering \\results.} 
\label{fig:partition-clusters}
\end{subfigure}
\caption{Partitioning is necessary to detect adjacent tables.}
\label{fig:partitioning}
\end{figure}

However, considering connected components as elements could lead to incorrect region boundaries:
as highlighted by Figure~\ref{fig:partition-csv}, sometimes regions can be adjacent to each other.
In the example, different rectangular regions compose a single connected component with irregular edges (Figure~\ref{fig:partition-coarse}).
Therefore, to identify a valid set of elements that leads to correct region boundaries, we need a segmentation strategy for connected components.
We ``cut'' the connected components along their non-concave edges (Figure~\ref{fig:partition-cut}).

Formally, we partition the components following a \emph{rectilinear} cut that is obtained by extending the edges incident to concave vertices towards the interior of the polygon, until a polygon boundary is met. 
Bajuelo et al.\ show that each given polygon, with $v$ concave vertices, can be split into $O(v^2)$ elements, with $2v+1$ as a minimum~\cite{bajuelos2004partitioning}.

With this method, even coherent elements could be initially decomposed.
This is eventually corrected while searching for regions in the next phase -- clustering -- where finer-grained elements can be either merged or not, granting the ability to even discover regions that appear directly adjacent in the spreadsheet (Figure~\ref{fig:partition-clusters}).

\subsection{Region detection and matching}
\label{subsec:region-detection}

The next phase of \systemname has the objective of clustering together elements that belong to the same region.
For a given spreadsheet, we have no prior knowledge of the number of regions that it contains.
Thus, we cannot use \emph{centroid-based} clustering approaches, such as k-means.
Instead, we resort to a customized \emph{density-based} approach, modifying DBSCAN~\cite{dbscan1996} to operate with a custom distance metric that highlights the structural properties we seek.
The DBSCAN optimization problem aims at finding points in dense neighborhoods of a given space: if we consider spreadsheet elements as points, a region corresponds to an area with a high density of points.
Given a distance function and a minimum number of points $m$ that form a cluster, the algorithm defines as \textit{core points} of a cluster all those that have at least $m$ points closer than a threshold $\varepsilon$, also called the \textit{radius} of the search space. 
Then, it groups all points that are within~$\varepsilon$ from a core point, or within $\varepsilon$ from non-core points belonging to a cluster.

In the original DBSCAN algorithm, every leftover point is labeled as noise. 
In our scenario, we are interested in labeling all the elements of a spreadsheet. 
Therefore, we do not consider any element as noise and set the minimum number of elements that can form a region as $m=1$. 
The distance function we use to compare elements is a weighted sum of three terms:
\begin{enumerate}
\item \textbf{Distance}: The Euclidean distance of their closest cells (Definition~\ref{def:distance}).
\item \textbf{Size  difference}: Considering $a_0, a_1$ as the areas of two elements, with the larger being $a_1$, the ratio $ 1 - a_0/a_1 $. 
\item \textbf{Alignment magnitude}: The number of shared points across the horizontal or vertical axis (Definition~\ref{def:alignment-magnitude}).
\PaperLong{
Defining $(x^i_{\textit{TL}}, y^i_{\textit{TL}})$ as the coordinates of the top-left corner of an element $i$ and conversely $(x^i_{\textit{BR}}, y^i_{\textit{BR}})$ as those of the bottom-right corner, the alignment is calculated as the sum of horizontal alignment, $h = |y^0_{\textit{TL}} -y^1_{\textit{TL}}| + |y^0_{\textit{BR}} - y^1_{\textit{BR}}|$ and vertical alignment $v = |x^0_{\textit{TL}} -x^1_{\textit{TL}}| + |x^0_{\textit{BR}} - x^1_{\textit{BR}}|$.
}
\end{enumerate}

\PaperLong{
In the first term, we decide to take the distance between the closest cell of two elements to avoid the influence of element size in the calculation. 
In facts, even if two elements are adjacent, if their width/height extends much farther than the boundary they share, any other distance metric (e.g., the distance of their center points) would be dependent on the width/heights of the elements rather than on their visual closeness.
order to reflect the geometrical distance between two elements even in the case  larger elements.

The intuition behind the second term, which is inversely proportional to the difference in size of the two elements, is that on one hand, two elements that are equally small or large, such as two metadata regions or two tables, are more probable to be two independent regions; on the other hand, larger elements are more likely to be grouped with smaller elements, such as tables and footnote regions.

The alignment is calculated as the sum of horizontal alignment and vertical alignment. This term is meant to compensate the effect of empty cells within regions: if different elements that are separated by visual space have a high alignment, they are most likely to belong together, e.g., like two parts of a table that are separated by an empty column.
}
    
The weights for these terms are $\alpha, \beta, \gamma$, respectively, and can be fine-tuned globally or for a given spreadsheet as hyperparameters for optimal boundary detection.
Additionally, the value of the radius $\varepsilon$ plays an important role in the success of the clustering, as different files can have different properties regarding the size of regions and the mutual distances of their elements.
We hypothesize that larger spreadsheets have, on average, a higher number of elements with greater distances, and therefore benefit from larger radii.
As Section~\ref{subsec:region-detection} points out, the best performances are obtained when setting a custom radius for each file. 
To reflect a scenario with no specific hyperparameter selection, we also experimented with our approach to find a suitable fixed hyperparameter setting for all files. 



Once their boundaries have been identified, we are interested in equivalent regions.
Our definition for region equivalence (Definition~\ref{def:region-equivalence}) is based on element boundaries and their types: for example, two footnote regions are equivalent if their entire content is equal, while two tabular regions are equivalent if their header elements are the same, regardless of the actual data content.

As Mondrian lacks semantic knowledge about cell types and relies on image segmentation and clustering to identify element and region boundaries, we need a suitable similarity measure to estimate actual equivalence.
\change{Moreover, due to its complexity, we do not compute graph isomorphism for region matching but rather compute region similarity based on syntactic cell types and their color encoding.}
Note from Table~\ref{tab:datatypes} how our color encoding assigns one primary color (red, green, blue, white) to each fundamental data type and then varying shades of the primary color to each sub-type belonging to the same fundamental data type.
For example, \textit{string} is associated with red, with \textit{lowercase} being ``tomato red'' (RGB $(255,75,75)$) and \textit{titlecase} being ``scarlet red'' (RGB  $(255,0,0)$).

In this way, cells with the same fundamental data type but different sub-types are more similar in the color space than cells from different fundamental types.
A given region is described with the color histograms of its cells, computed with 64~bins for each channel, for a total of 192~bins.
The color histogram is a global descriptor of each region that acts as a region ``fingerprint'': its values are dependent on the amount and distribution of cells of different types.
The similarity of any two regions is then computed as the cross-correlation of their color histograms.
Furthermore, the color encoding can be easily extended including more, or further refined, data types.
If two highly similar regions (that is, whose similarity is over a given threshold) are found in two different files, they are considered equivalent and the file layouts that contain them are candidate instances of the same template (cf.\ Section~\ref{subsec:template-inference} for a detailed explanation).

\subsection{Layout similarity}
\label{subsec:layout-similarity}

Each spreadsheet file, once its regions have been detected, has an associated file layout, represented as a complete graph with regions as nodes and labeled edges that describe their spatial relationships (Definition~\ref{def:file-layout}).
\change{
As with region equivalence, we do not compute exact graph isomorphisms for layout equivalence but rather approximate it with a similarity measure.
}
Our algorithm is based on the similarity flooding approach proposed by Melnik et al.\ for graph matching~\cite{melnikGR02}.
The core intuition is to first compute an initial pair-wise similarity of nodes across the two file layout graphs using the region similarity metric described in Section~\ref{subsec:region-detection}.
If the graph $\mathcal{G}_a$ has $U$ nodes and the graph $\mathcal{G}_b$ has $V$ nodes, we obtain a matrix $\sigma^0$ of $U\times V$ values.

Additionally, we build a $\binom{u+1}{2}\times\binom{v+1}{2}$ matrix $\Phi$ of edge similarities, where the value in position $\Phi(i+j,\ k+l)$ with $i,j,k,l \in \mathbb{N}_0$ corresponds to the edge similarity of $\textit{edge}(u_i,\ u_k)$ and $\textit{edge}(v_j,\ v_l)$. 

The edge similarity is set to 0 if any of the node pairs $(u_i,\ u_k) \in \mathcal{G}_a$ , $(v_j,\ v_l) \in \mathcal{G}_b$ have no connecting edge (including the case of both being the same node), or if the two edges have a different \textit{alignment direction}.
Otherwise, the edge similarity is computed as the Euclidean distance between the vectors composed of the features \textit{(alignment magnitude, distance)}, normalized by the maximum value to have a similarity score in $[0,1]$.

The similarity of the nodes in $\sigma^0$ is then iteratively ``flooded'' by multiplying the similarity of each node pair with the similarity of the neighboring node pairs, weighted by the edge similarity in $\Phi$.
In formal terms, the similarity of the $i$-th node of $\mathcal{G}_a$ and the $j$-th node of $\mathcal{G}_b$ is iteratively updated using the formula 
$$\sigma^{k}(i,j) = \sigma^0(i,j)+\sum_{\substack{m =0...V,\ n = 0...U}}{\sigma^{k-1}(m,n)\cdot\Phi(i+m, j+n)}$$

As we look for a 1:1 node match, we ensure that for every neighboring node pair $(u_i,\ u_j) \in \mathcal{G}_a$, only the node pair $(v_j,\ v_l) \in \mathcal{G}_b$ with the maximum edge similarity is used.
To avoid imbalance in similarities for node pairs ($u$,$v$) where any of $u$ or $v$ has a high number of neighbors, we normalize the value of $\Phi$ dividing $\Phi(u+v,u_i+v_j)$ by $2^{n-m}$, where $n,\ m$ are the number of neighbors of $u$ and $v$, respectively.
Finally, at each iteration, we normalize the values of $\sigma^i$.
The iterative computation is stopped either when the matrix distance $||\sigma^{i+1},\sigma^i||_2$ falls below a given threshold, or when a maximum number of iterations is reached.
During our experimentation, we empirically observed that in most cases the matrix difference falls quickly (in a handful of iterations) to values in the range $[0.01,0.1]$ and then stabilizes, reaching values under 0.01 with a much slower convergence speed (in thousands of iterations).
Therefore, we recommend setting a threshold of 0.1 and a maximum number of iterations to 10, which we deem sufficient considering the satisfactory results obtained on the template inference task reported in Section~\ref{subsec:evaluation-template}.

At the end of the similarity flooding stage, we can consider the matrix $\sigma$ as the weight matrix of a fully connected bipartite graph $\mathcal{B}$, with the two partitions composed of the nodes of $\mathcal{G}_a$ and $\mathcal{G}_b$, respectively.
To compute the final similarity score of ($\mathcal{G}_a$, $\mathcal{G}_b$), we find a maximum weighted matching on $\mathcal{B}$ and average the corresponding weights found, including zero values in the computation for every $|\ |\mathcal{G}_0$| — |$\mathcal{G}_1$|\ | node left unmatched.
In formal terms, given the weights $w(u,v)$ for nodes $u\in\mathcal{G}_a, v\in\mathcal{G}_b$, the similarity between $\mathcal{G}_a$ and $\mathcal{G}_b$ is computed as:
\[
    sim(\mathcal{G}_a, \mathcal{G}_b) = \dfrac{\sum\limits_{u \in \mathcal{G}_a , v \in \mathcal{G}_b}{w(u,v)}}{max(|\mathcal{G}_a|, |\mathcal{G}_b|)}
\]

As this graph similarity is asymmetrical, because of the matrix normalization included in the calculations, for every pair of files $f_a, f_b$ we compute the final file layout similarity $sim(f_a,f_b)$ averaging between $sim(\mathcal{G}_a,\mathcal{G}_b)$ and $sim(\mathcal{G}_b,\mathcal{G}_a)$.
\PaperLong{
As proven by Melnik et al., in the case of fully connected graphs, the layout similarity computation has a complexity of $O(u^2 \cdot v^2)$ for two files with $u$ and $v$ regions~\cite{melnikGR02extended}.}

\subsection{Template inference}
\label{subsec:template-inference}

The ultimate goal of \systemname is to find spreadsheet layout templates. 
As we approximate pairwise layout equivalence with our graph-based similarity measure, we consider two files layout to be instances of the same template if their pairwise similarity is above a given threshold $\tau_f$ (subject to evaluation in Section~\ref{subsec:evaluation-template}).

To extend template inference beyond pairs of files, we use an ``inductive'' approach: given a set of files, each with its detected regions, we examine the set iteratively.
\PaperLong{
The procedure for template recognition is described in pseudocode in Algorithm~\ref{alg:mondrian}, in lines 8-24.
}
The first file $f_0$ is considered an instance of a template $t_0$, and its regions are added to a global index of regions $\mathcal{R}$, along with the information that these regions are found in the layout of $f_0$\PaperLong{~(Line 17)}.
When a new file $f$ is examined, first its regions are compared with all the regions in $\mathcal{R}$\PaperLong{~(Line 10)}. 
If a region $r_{f}$ is similar to a region $r_{t}$ in $\mathcal{R}$ more than a threshold $\tau_r$, we add the file layouts that contain $r_{t}$ to the list of possible similar layout candidates for the file $f$\PaperLong{~(Line 12)}.
\change{
If no region in $\mathcal{R}$ matches any of the regions in $f$, \systemname will not compute any pairwise layout similarity.}
During our experimentation, we discovered a region threshold $\tau_r=0.75$ to be sufficient to obtain valid similar layout candidates.
If the layout of the file $f$ has a similarity greater than $\tau_f$ to the layout of a candidate file $f_t$, we group $f$, $f_t$, and, recursively, all files grouped with both $f$ and $f_t$.
\PaperLong{
We do so by creating a graph with a node for each file, and an edge connecting two nodes if the layout similarity is above the layout threshold $\tau_f$ (Lines 19-22).
Templates are found as the connected components of such graph~(Line 23).
}
In this way, we assume templates are transitively closed.
Nonetheless, the results for a file set are independent of the order the spreadsheets are processed: at the last iteration, all regions will have been compared against each other, and so will all layouts containing matching regions. If at any given point a file is found matching two distinct templates, these are merged.
We choose this iterative approach for different reasons: first, it suits a continuous development scenario, where the region index and template layouts are persistently stored and can be reused in later stages as new files are pre-processed.
Second, it is significantly less computationally expensive to pre-compute region similarities and prune the template search space rather than perform graph similarity for each pair of files, which would anyway include computing the pairwise region similarity for all pairs of regions found across all files.
\PaperShort{In the extended version of this paper, we provide a pseudocode for the end-to-end \systemname approach and a discussion of its theoretical complexity \cite{extendedMondrian}.
}

\PaperLong{
\begin{algorithm}
\begin{algorithmic}[1]
\State Input: set of input files $\mathcal{F}$, clustering parameters $\alpha, \beta, \gamma$, region similarity threshold $\tau_{r}$, layout similarity threshold $\tau_f$
\State Output: templates $\mathcal{T}$, regions $\mathcal{R}$
\State $\mathcal{R} \gets \{\}$ \# $\mathcal{R}$ region dictionary
\State $\mathcal{T} \gets \{\}$ \# set of templates
\State $P \gets \varnothing$ \# set of candidate file pairs
\For{ $f$ in $\mathcal{F}$ }
    \State img $\gets$ parse($f$)
    \State $R_f \gets$ region\_detection(img, $\alpha, \beta, \gamma$)
    \For {$r_{f}$ in $R_{f}$}
        \For {$r_t$ in $\mathcal{R}$.keys}
            \State $\sigma_r \gets$ region\_similarity($r_{f}$, $r_t$)
            \If { $\sigma_r \geq \tau_r $}
                \State $P \gets P \cup  \{ (f,f_t)$ for $f_t \in \mathcal{R}\langle r_t \rangle \}$ 
                \State $\mathcal{R}\langle r_t \rangle \gets \mathcal{R}\langle r_t \rangle \cup \{f\} $
            \Else
                \State $\mathcal{R}\langle r_f \rangle \gets \{f\}$
            \EndIf
        \EndFor
    \EndFor
    \If {$|\mathcal{R}| == 0$}
        \State $\mathcal{R} \gets \{ r_f : \{f\}\ \forall\ r_f \in R_f \}$
    \EndIf
\EndFor

\State $G_s \gets \{f$ for $f \in \mathcal{F} \}$ \# similarity graph
\For {$(f,f_t)$ in $P$}
    \State $\sigma_f \gets$ layout\_similarity($f$, $f_t$)
    \If {$\sigma_f \geq \tau_f$}
        \State Add edge in $G_s\langle f, f_t \rangle$
    \EndIf
\EndFor
\State $\mathcal{T} \gets $ find\_connected\_components($G_s$)
\State \Return $\mathcal{T}$, $\mathcal{R}$
\end{algorithmic}
\caption{\change{Pseudocode for the Mondrian approach}}
\label{alg:mondrian}
\end{algorithm}



\subsection{Complexity Analysis}
\label{subsec:complexity}

Formally, the complexity of \systemname depends from three main procedures: the region detection, the region similiarity and the file layout similarity.
The region detection runs for each of the $F$ files in the dataset: for a file containing $E$ rectangular elements, 
DBSCAN has an average complexity of $O(E\cdot logE)$ and a worst case complexity of $(O(E^2)$ \cite{dbscan1996}.
In the worst case, $E$ is equal to the number of non-empty cells of a file.

The complexity of region and layout similarity depends on the number of files $F$, the number of overall regions across files in the dataset $M$, as well as the number of unique regions $N$.
The cost of computing parwise region similarity is constant, as it is a single operation on two fixed-length vectors, whereas the cost for pairwise layout similarity for two files containing $N_1$ and $N_2$ regions is $O(N_1^2 \cdot N_2^2)$, as reported by Melnik et al.\ ~\cite{melnikGR02extended}.

If no regions are similar, i.e., $M=N$, the region similarity stage has a cost of $O(M^2)$, but \systemname computes no layout similarity.
If all files contain the same regions , i.e., $M=N\cdot F$, the region similarity is computed $O(N^2 \cdot F) = O(M \cdot N)$ times. 

In this worst case scenario no pruning happens, and the complexity for layout similarity is $O(N^4 \cdot F^2)$.
Typically, the number of regions in a file is reasonably lower than the number of files: for example, \DECO and \FUSTE, the two real-world datasets used for our experiments described in Section~\ref{subsec:evaluation-dataset}, both contain above 800 files with on average 4.43 and 2.09 regions per file, respectively.
Therefore, assuming $N<<F$ the upper bound for complexity is given by $O(F^2)$.

Finally, the transitive closure operation to obtain templates has a complexity of $O(F \cdot S)$, where $S$ stands for the number of pairwise similar files: in the worst case scenario, $S=\frac{F(F-1)}{2}$ and the complexity is $O(F^3)$.

In our empirical observations, the average complexity of template recognition for \systemname behaves quadratically with the number of files, as reported in Section~\ref{subsec:evaluation-template}.

}

\section{Evaluation}
\label{sec:evaluation}
\Multifile spreadsheets pose interesting data engineering challenges.
In Section~\ref{subsec:problem} we described three related research problems: region detection, region matching, and template inference.
We conducted a set of experiments to evaluate whether it is possible to address these problems using an automated approach that is general with respect to the spreadsheet format, 
and with respect to domain knowledge.
We compare \systemname to a system that uses connected components to discover tables~\cite{coletta_public_2012}, an approach for genetic algorithm-based table recognition~\cite{koci2019genetic}, and a CNN-based machine learning model~\cite{dongLHFZ19}.


\subsection{Evaluation datasets and their properties}
\label{subsec:evaluation-dataset}

To evaluate our approach, we use two datasets of real-world spreadsheets.
The first, \DECO~\cite{koci2019deco}, is a publicly available annotated file sample of enterprise spreadsheets extracted from the ENRON corpus~\cite{enron2015}.
It is composed of 1,165 MS~Excel files used in an energy company and found in email attachments from 2000 to 2001, annotated by Koci et al.~\cite{koci2019deco}.
Of those, roughly 27\% are classified by the authors as not containing a table (e.g., containing only charts).
For the remaining 854 files, in the case of multiple worksheets per file, the authors annotated only one worksheet with regions.
We use these regions as candidates for our region detection task.
In addition, we manually annotated the dataset at the file level to identify files with the same layout, for the template inference task\footnote{Available at 
\url{\vldbavailabilityurl}}.

The second dataset is sampled from FUSE, a large-scale corpus of spreadsheets crawled from various internet sources~\cite{fuse2015}.
For our evaluation, we annotated the region layout and the templates of all relevant 886 worksheets from 780 unique, randomly sampled spreadsheet files.
In the remainder of this section, we call this annotated subset \FUSTE (FUSE Sample for Template Extraction).
The region-level annotations of \FUSTE have been obtained with the tools proposed in the original \DECO paper~\cite{koci2019deco}, to stay consistent with those from this dataset.
Table~\ref{tab:datasets} reports the main characteristics of the two datasets concerning their files' layouts.
The first consideration is the wide presence, in both sources, of \multifile files: roughly 72\% and 45\% of files from \DECO and \FUSTE, respectively, have more than one region.
\begin{table}[t]\small
\begin{tabular}{rrr} & \multicolumn{1}{c}{\textbf{\DECO}} & \multicolumn{1}{c}{\textbf{\FUSTE}} \\
\toprule
Total number of files & 854 & 886 \\
Files with one/multiple regions & 233/621 & 495/391  \\
Overall layout templates & 750 & 136 \\
Templates with one/more than one files & 679/71 & 105/31 \\
\bottomrule
\end{tabular}
\caption{Synthetic overview of the evaluation datasets.}
\label{tab:datasets}
\vspace{-20pt}
\end{table}
\FUSTE has overall a greater number of single region files and on average much fewer regions per file than \DECO (2.09 and 4.43, respectively), with \DECO having more files with a huge number of regions -- the maximum being 321. 
For the rest of the experiments, we regard as outliers, and therefore exclude, those files with more regions than the 99.9\% of the remaining files in the same dataset. 
These files, two for \DECO and one for \FUSTE, were characterized by an unusually large number of regions sparsely distributed across the spreadsheet. 
The two datasets also show opposite natures regarding layout templates.
\DECO has a low level of layout recurrence, with 750 different layout templates for 854 files, 679 of which are ``singletons'', i.e., covering only one file.
\FUSTE, on the other hand, contains 136 templates for 886 files, with one encompassing as many as 381 different files and only 105 singleton templates. \systemname handles both extremes well. 

\PaperLong{
Various considerations arise from these fundamental differences.
First, both manually annotated datasets represent a relatively small sample of the entire collection of files from the original corpora: ENRON, the source of \DECO, is composed of 15,770 unique spreadsheet files with 79,983 sheets~\cite{enron2015}; FUSE, the source of \FUSTE, has 249,376 unique spreadsheets.
We are confident that the templates we discover typically cover many more spreadsheets than in our sample.
Additionally, the origin and thus usage of the two datasets are different: \DECO is a set of enterprise spreadsheets, in which files have possibly a ``single-use'' scope, be it for reporting or analysis purposes within the company; \FUSTE is a set of documents crawled from various public internet sources, most likely designed for sharing, with a high homogeneity of files originating from the same source.
Finally, as surfaced during our annotation of \DECO templates, this dataset could have shown a higher percentage of file similarity with a different choice of the worksheets:
the choice of annotating only one worksheet per file excluded various worksheets from the original files that showed the same layout.
}
\subsection{Related approaches for comparison}
\label{subsec:baselines}

The experiments conducted to evaluate the performance of our region detection approach include, for comparison, the results obtained on the same task using the connected component detection algorithm outlined in the work of Coletta et al.~\cite{coletta_public_2012}, the genetic-based table recognition approach proposed by Koci et al.~\cite{koci2019genetic}, and the CNN-based TableSense~\cite{dongLHFZ19}.
Furthermore, simply selecting Coletta et al.'s connected component approach can be considered a baseline for our approach: it is the first step from which we build upon element partitioning and clustering.

The genetic-based approach is a more sophisticated process, involving two steps that rely on supervised machine learning methods.
In the first step, a random forest classifier is trained on cell features to label each spreadsheet cell according to its role (e.g., \textsf{data}, \textsf{header}, \textsf{aggregate})~\cite{koci2016machine}.
Afterward, neighboring cells with the same label are grouped and a graph is formed, with cell groups as vertices and their spatial relationship as edges~\cite{koci2018graph}.
Different tables are recognized as sets of vertices obtained by partitioning the graph~\cite{koci2019genetic} using a supervised genetic-based algorithm.
This overall approach relies on rich features extracted from Excel files and aims at solving the more complex task of table recognition.
Recall that the region detection task we solve is slightly different in goal and assumptions: we are interested in detecting region boundaries in general \multifile~spreadsheets, without assuming special formatting features nor tabular structures.

The comparison was conducted with the help of the original authors, 
reusing the source code for the feature extraction, cell classification, and the genetic approach\footnote{\url{https://github.com/ddenron/gen\_table\_rec}, last accessed Feb 25, 2020}.
For a fair comparison, we experimented with two versions of the genetic-based approach: one using the full set of Excel-specific features available, and one restricting the input information to only cell content and position, excluding style features, thus simulating a \emph{.csv} file input.
The model, following the setup described by the authors in~\cite{koci2019genetic}, is trained and tested on each dataset using 10-fold cross-validation.

TableSense, proposed by Dong et al.~\cite{dongLHFZ19}, is based on Mask R-CNN~\cite{heGDG17MaskRCNN}, a convolutional neural network developed for instance segmentation in images.
TableSense extends this architecture for the task of table detection in spreadsheets with two specialized modules: a feature extraction stage to map spreadsheets into feature maps that are served as input to the network, and a Precise Bounding Box Regression layer to refine the coordinates of Mask R-CNN detected regions' bounding boxes.
The intuition of TableSense, like \systemname, is to map the region detection task to the visual domain: using a convolutional architecture, it leverages the 2D distribution of cells on a spreadsheet to identify ``Regions of Interest'', candidate areas of the input file, which are then classified as tables and whose boundaries are refined by the PBR module.
The authors report experimental results of TableSense training the model on the WebSheet10K dataset and testing it on the WebSheet400 dataset.
As neither the trained models nor the original source code is publicly available, to compare it with \systemname in a similar setup we obtained the results training the model on one dataset and testing on the other, i.e., the results for \DECO are obtained training TableSense on \FUSTE and vice-versa.
Due to the non-deterministic nature of the approaches that involve machine learning approaches (Genetic-based and TableSense), we repeated the experiments involving the full pipeline three times, and report average scores, with confidence intervals obtained from the standard deviation of the experiment results.

For the region detection stage of \systemname, we use two setups regarding the choice of the clustering radius: one using an optimal, ``dynamic'' choice of the clustering radius for each file, and one with a ``static'' radius used across all dataset files.
In the dynamic radius setting, we ran our clustering method on each file, varying the size of the radius between [0.1,2] in steps of~0.1, between [2,10] in steps of~1; and between [10,100] in steps of~10.
Additionally, we experimented with different configurations of the distance features' weights:
we kept $\alpha=1$ as a fixed reference value and varied $\beta, \gamma \in \{0, 0.5, 1, 5, 10\}$.
The hyperparameter configuration that showed the best results was $\alpha=1, \beta=0.5, \gamma=1$ for \DECO, and $\alpha=1, \beta=1, \gamma=1$ for \FUSTE.
We use these values for experimenting in the ``static'' radius setting, in which we tried to find the single radius that showed the best performances across all files. 
The search space for the radii was the same as the one used in the dynamic setting.
We report the result obtained using the radius with the best performance for each dataset, namely~1.5 for \DECO and~1.4 for \FUSTE.
\subsection{Region detection accuracy}
\label{subsec:evaluation-iou}
To evaluate the level of accuracy in region detection, we use the Intersection-over-Union score (IoU)\PaperShort{, the graphical equivalent of the Jaccard index for sets\footnote{In the extended version of this paper \cite{extendedMondrian}, we also report the performances obtained with the EoB score, an additional region similarity metric proposed in~\cite{dongLHFZ19}. These results point to the same outcomes.}.} 
\PaperLong{ and the Error-of-Boundary score (EoB), defined in~\cite{dongLHFZ19}.
The first value is the graphical equivalent of the Jaccard index for sets.}
\change{If we define $P$ as the set of non-empty cells of a predicted region, and $T$ as the set of non-empty cells of a target region, the IoU is calculated as:
$$ IoU(P,T) = \frac{|P \cap T|}{|P| + |T| - |P\cap T|}  $$
}
\PaperLong{
If we define a region's top-left coordinates as $(x_0, y_0)$ and bottom-right coordinates as $(x_1, y_1)$, for two regions $P$ and $T$ the EoB is calculated as:
$$ EoB(P,T) = \max(|P_{x_0}-T_{x_0}|, |P_{y_0}, T_{y_0}|, |P_{x_1}, T_{x_1}|, |P_{y_1}, T_{y_1}|)$$
}

An IoU score of~1 corresponds to perfectly detected regions and a score of~0 to missed regions.
\PaperLong{A perfectly detected region has an EoB of~0, with no upper limit for incorrect detections.
EoB is undefined in the case of no detected region: whenever such a case arises, we set the EoB as the maximum of the height and width of the file, simulating a completely out of boundary detection.}
\change{
The standard in literature is to and consider ``correctly detected'' all true regions for which the score of at least one predicted region exceeds a given threshold
\cite{koci2019genetic,gilani2017table,dongLHFZ19}.
To provide more accurate results, we measure actual scores rather than their binarization.
In general, any of the true regions $R_T$ of a file can be split into multiple $R_P$ predicted regions, or vice-versa, one of the predicted regions can span multiple true regions.
Therefore, for $M$ predicted regions and $N$ true regions IoU determines $M \cdot N$ scores: to achieve only one value for a given true region, we assign it to the predicted region with the highest overlap:
$$ IoU(T) = \max_{P \in R_P} IoU(P,T)$$
}
\PaperLong{$$ EoB(T) = \min_{P \in R_P} EoB(P,T)$$}

\PaperShortFigure{
\begin{figure}[t]
\centering
\includegraphics[width=\columnwidth]{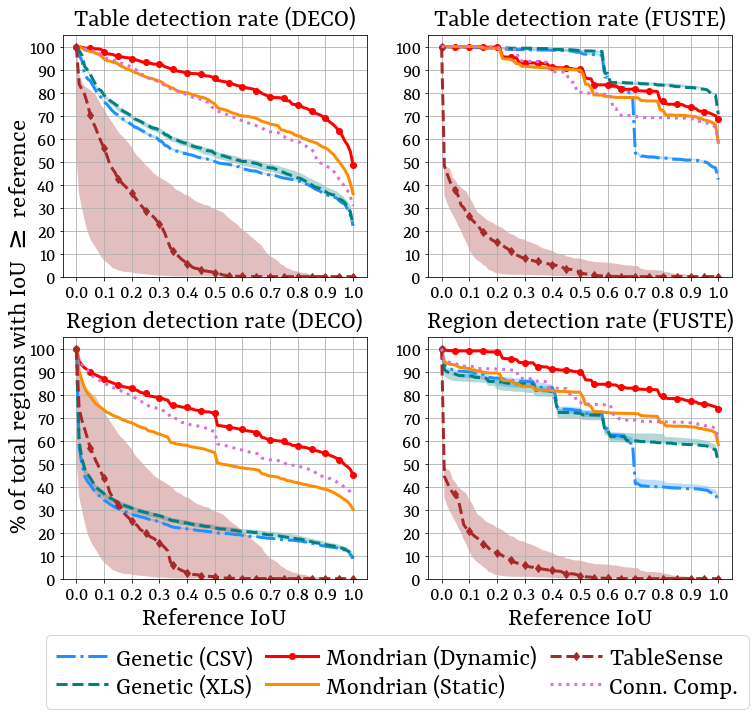}
\caption{Table and region detection performance.}
\label{fig:iou-comparison}
\vspace{-10pt}
\end{figure}
}
\PaperLongFigure{
\begin{figure*}[t]
\centering
\includegraphics[width=\textwidth]{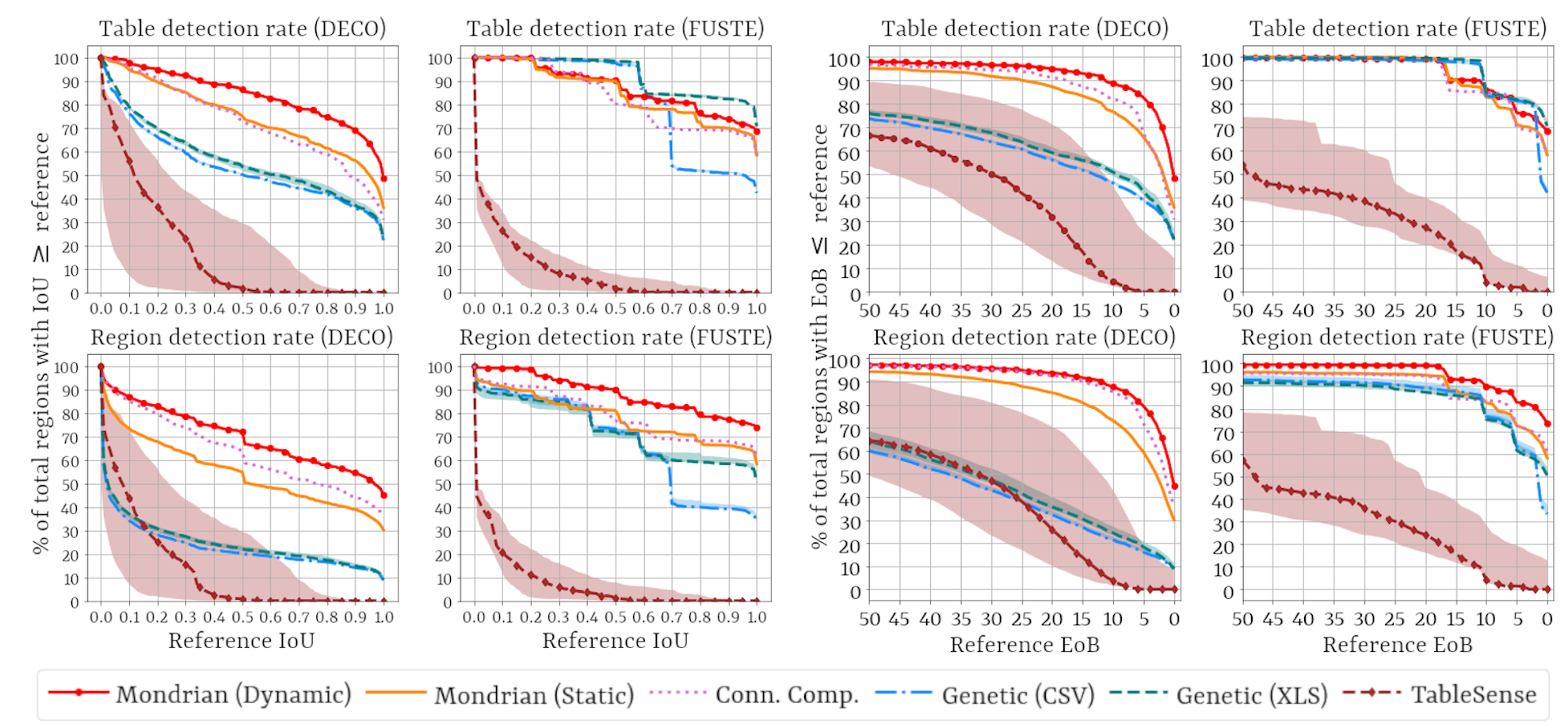}
\caption{Table and region detection performance.}
\label{fig:iou-comparison}
\end{figure*}
}

Figure~\ref{fig:iou-comparison} shows the performance of the different approaches over varying thresholds: the y-axis represents the percentage of tables or regions correctly detected in the two datasets, assuming as ``correct'' a score better than the given reference on the x-axis.
We report the performance for tabular regions only (``table detection''), and the performance across all types of regions (``region detection''), which include tables but also notes, spreadsheet titles, etc.

\subsubsection{\change{\systemname performances}}
The best results for all regions are obtained, for both datasets, with our clustering approach assuming a dynamic, optimal choice of the radius for each file.
It is interesting to note the difference in the behavior of \systemname on the two different datasets. 
\DECO, which contains more \multifile files and on average more regions per file, proves to be the harder of the two with approximately 45\% of regions perfectly detected (100\% IoU).
On FUSE, instead, with fewer complex multiregion files, around 75\% of the regions are correctly detected. 
The usage of a static radius yields lower performance: in the case of tables the accuracy is comparable to detecting connected components, while on other region types it yields slightly worse results.
In our experiments, a smaller radius ($\leq 1$) made the clustering degenerate into connected component detection, grouping only adjacent partitioned elements.
A larger radius, such as the one selected for our static approach (namely 1.5), improves table detection, since a high number of tables is composed of separated connected components, but also brings together different non-tabular regions, which are usually independent.
Because of this, the static radius variant of our clustering approach shows slightly worse performance in detecting general regions than tables.

\subsubsection{\change{Comparison with the genetic-based approach}}
\label{subsec:results-generic}
It is not surprising that the genetic-based approach shows better results for tables than for generic regions, as it was specifically designed for table recognition.
When cell classification and table detection are combined end-to-end, the second step proved to be sensitive to even small errors in the cell classification, with the results visible for the \DECO dataset in Figure~\ref{fig:iou-comparison}.
On \FUSTE, where the classification errors were minimal, the genetic-based approach showed much better results.
We explain this phenomenon by considering the reliance of the genetic-based search on correctly labeled region boundaries.
The incorrect classification of some cells causes the split of one single region into different vertices, some of them necessarily erroneous.
Moreover, it appears that non-data cells, such as header or aggregation cells, are crucial for recognizing tabular structure.
Such classification errors propagate into unreliable weight learning for the quality measures of the fitness function, and finally cascade into poor table boundaries.
It is worth noting how, for \FUSTE, the contribution of Excel-specific features is much more significant than for \DECO: the gap between the two versions of the genetic approach is much wider.

\subsubsection{\change{Comparison with TableSense}}
\label{subsec:results-tablesense}
The results of TableSense show low performance with a high variance. 
This behavior can be explained by noting the considerable number of regions that are completely missed: on average, 48.81\% for \DECO and 32.92\% for \FUSTE.
Contrarily to \systemname, which by design does not ignore any non-empty input cell, the CNN architecture of TableSense may completely ignore entire areas of the input if they are not considered ``Regions of Interest'' or classified as containing an object.
This behavior is inherited from the original domain of Mask R-CNN, designed for instance segmentation of images, which may or may not contain relevant objects.
Overall, the poor accuracy of TableSense is most likely due to the high complexity of the model, which is composed of more than 85 million trainable parameters, and the limited number of training files available for our use case.

\PaperLong{
\smallskip
Comparing the plots across the two datasets, we note an interesting difference: the \DECO plots are much smoother than those of \FUSTE, which show abrupt drops in the percentage of tables and regions recognized above a certain threshold.
This phenomenon reflects the different dataset natures, as analyzed in Section~\ref{subsec:evaluation-dataset}.
Considering that \DECO has roughly twice the quantity of regions compared to \FUSTE, it is natural to expect a more continuous plot.
What is more, \FUSTE contains a greater number of files that share the same templates: on average, 5.33 files sharing the same layout compared to the 1.13 in \DECO.
In particular, one can observe how the percentage of tables (and regions) detected correctly in \FUSTE drops from 80\% (60\%) to 50\% (40\%) for the Genetic-CSV approach as soon as the threshold for the IoU is increased from 69\% to 70\%.
Looking for the causes of this behavior, we found that 323 different regions, coming from just as many files with the same layout, were detected in the same way.
The absence of the same drop from the Genetic-XLS approach suggests that including style features helped in recognizing these regions correctly.
}

\change{
\subsubsection{Sensitivity to region composition}

\begin{figure}[t]
\centering
\includegraphics[width=\linewidth]{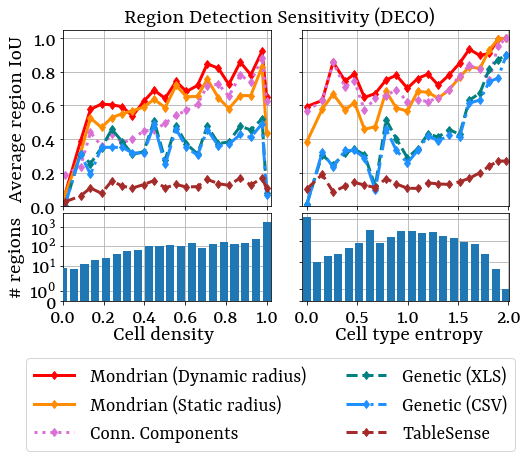}
\caption{\change{Performances per region composition.}}
\label{fig:region-density}
\vspace{-10pt}
\end{figure}

Considering the graphical nature of the clustering performed by \systemname, its performance on region detection is sensitive to the visual composition of regions.
To provide insights into the behavior of the different region detection strategies, we analyzed the effect of two variables: 
the density of a region, i.e., the ratio of non-empty cells to empty cells contained in a region, and the cell type entropy, i.e., the entropy of a region, which we calculate as $-\sum_{i=1}^{k}P(c_i)\cdot \log{P(c_i)}$, with $P(c_i)$ being the ratio of cells of (syntactic) type $i$ over the total cells of a region.
Figure~\ref{fig:region-density} reports the average IoU scores of the regions of the \DECO dataset sorted by their density and entropy.
Both plots show that \systemname is most successful with visually heterogeneous regions: its performance increases with increasing cell type entropy and has a sharp drop for regions with either very low densities, signaling a high number of empty cells, or a low cell type entropy, where it is unable to perform its partitioning.
We note that the low score for regions with a density of 1, i.e., with no empty cells, is highly correlated to the score for an entropy of 0, as \numprint{1192} out of the total \numprint{3462} regions have both a density of 1 and an entropy of 0.
This behavior reflects the inefficiency of visual partitioning for regions with few ``visual irregularities''.
In facts, these regions are those where the connected component baseline outperforms \systemname.
}

\subsection{Template inference accuracy}
\label{subsec:evaluation-template}

In evaluating the template inference task, we rely on three external 
measures for clustering: \emph{homogeneity}, \emph{completeness}, and \emph{v-measure}~\cite{vmeasure07}.
The value range of all three scores is [0,1], with 1 being a perfect result.
Using the gold standard, homogeneity quantifies how many data points in each predicted cluster belong to the same template.
For our problem, in a perfectly homogeneous solution, all files that are grouped indeed share the same layout. 
Completeness, conversely, quantifies the percentage of elements from the same template that are grouped. 
V-measure
is the harmonic mean of homogeneity and completeness.
As described in Section~\ref{subsec:template-inference}, we group files transitively based on their layout similarity being above a given threshold.
We experimented with thresholds in the range [0.7,1] with a spacing of~0.01.
\PaperLong{
To save computational time while repeating the experiments for different thresholds, we did not calculate the layout similarities of pairs for which we can guarantee a threshold lower than~0.7.
This pruning was possible given the nature of our approach, where the similarity of two graphs is bound by the absolute difference in their number of nodes, normalized by the maximum number of nodes across the two graphs.
}
\subsubsection{\change{Effect of layout similarity threshold}}
\begin{figure}[t]
\centering
\includegraphics[width=\linewidth]{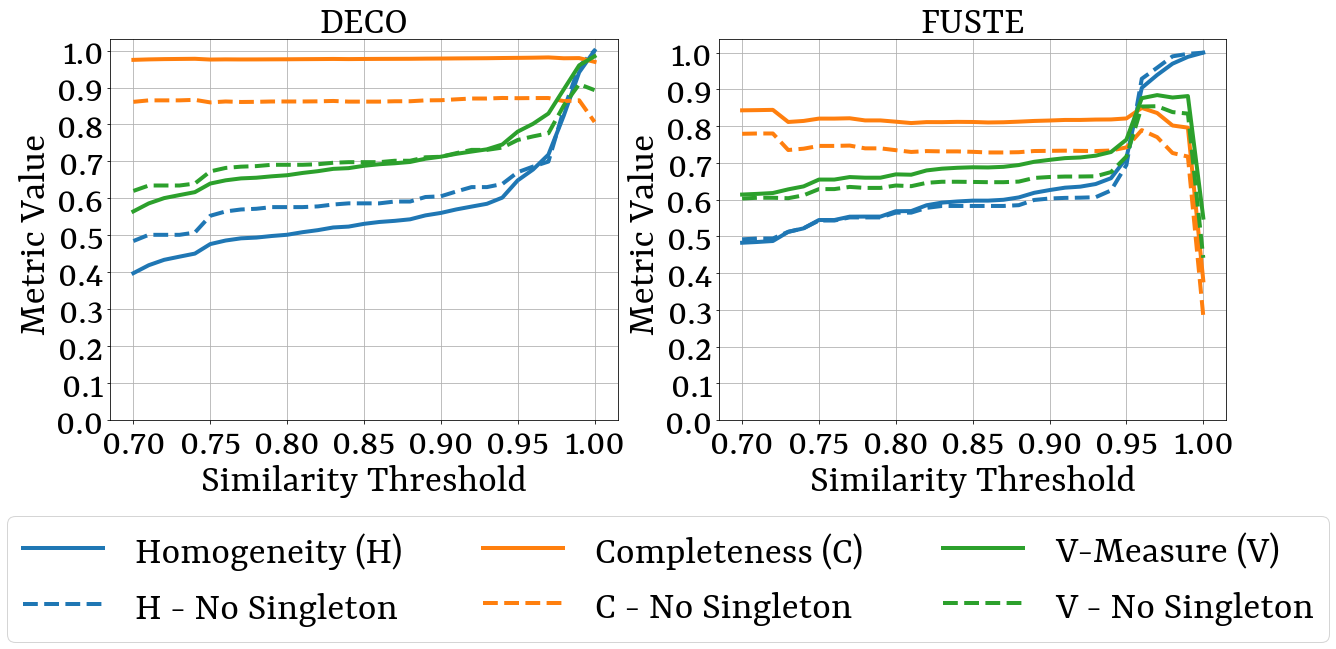}
\caption{Performance of \systemname on template inference.}
\label{fig:template-thresholds}
\end{figure}

Figure~\ref{fig:template-thresholds} shows the influence of the threshold value on the results of template recognition using the regions automatically detected by \systemname in the static radius scenario, for the \DECO and \FUSTE datasets.
Considering how, especially for \DECO, there is a significant number of singleton templates, i.e., templates that occur in only one file, we report the results of our template recognition approach for the full dataset as well as for the sub-set of files that constitute non-singleton templates (175 files for \DECO and 781 for \FUSTE, cf.\ Table~\ref{tab:datasets}).
Increasing the threshold leads to a more selective behavior: for the maximum threshold of 1, homogeneity reaches a perfect value, as the resulting templates are always comprised of one file and therefore trivially homogeneous.
This is compensated by the drop of completeness for high thresholds, especially noticeable in the \FUSTE dataset.
This effect is mitigated on the full \DECO dataset thanks to the high number of singleton templates.
Overall, the performances of our template inference approach benefit from choosing high thresholds: across the two datasets, the best v-measures are obtained with thresholds between 0.95 and 1.00.

\change{
\subsubsection{Sensitivity to number of regions}
To assess how the region composition of file layouts affect the template recognition performance, we partitioned the evaluation datasets in three groups: single region files, files with few regions (2 to 5), and files with many regions (more than~5). 
In Table~\ref{tab:templates-nregions} we report the scores obtained by \systemname on the three partitions using a threshold $\tau_f$ of 0.99.
Across both datasets the best performances are reached on files with large number of regions.
Conversely, the lowest homogeneity is obtained on single region files, where the layout graphs contain no edges (or presumably a few, due to errors in region detection).
This causes layout similarity to be mostly influenced by the approximate region similarity, that causes a slight increase of false positives.
}

\begin{table}[t]\small
\begin{tabular}{rrrrrrrrrr}
\textbf{Number of} & \multicolumn{4}{c}{\textbf{\DECO} ($\tau_f=0.99$)} & & \multicolumn{4}{c}{\textbf{\FUSTE ($\tau_f=0.99$)}}  \\
\textbf{regions}   & \#files & H & C & V & & \# files & H & C & V\\
\cmidrule[\heavyrulewidth]{2-5}\cmidrule[\heavyrulewidth]{7-10}
1 & 232 & 0.92 & 0.97 & 0.94 
& & 495 & 0.98 &0.68 & 0.80 \\
$[2,5]$ & 470 & 0.97 & 0.98& 0.98 
& & 372 & 0.98 &0.76 & 0.86 \\
$\geq 6$ & 150 & 0.99 & 0.98 & 0.99
& & 18 & 1.00 &0.95 & 0.97 \\
\cmidrule[\heavyrulewidth]{2-5}\cmidrule[\heavyrulewidth]{7-10}

\end{tabular}
\caption{\change{Template inference at varying number of regions.}}
\label{tab:templates-nregions}
\vspace{-2em}
\end{table}




\subsubsection{\change{Sensitivity to region detection strategy}}
The performance of our template inference algorithm is also dependent on the results of the prior region detection phase.
To analyze the sensitivity of the graph matching to region boundaries, we experimented with \change{all the region detection strategies considered in Section~\ref{subsec:evaluation-iou} plus a configuration using the manually annotated regions from the gold standard.}
\begin{figure}[t]
    \centering
    \includegraphics[width=\linewidth]{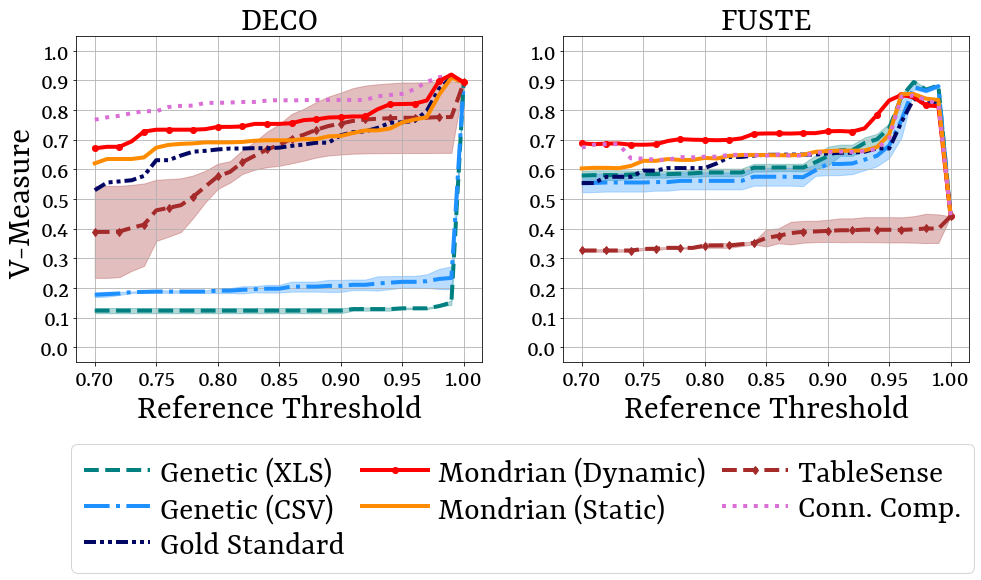}
    \caption{\change{Effect of region detection on template inference.}}
    \label{fig:template-scenarios}
\end{figure}
Figure~\ref{fig:template-scenarios} reports the v-measure for the different region detection strategies and baselines across datasets (excluding singleton templates).
First, we highlight how 
\change{approaches with poor region detection performances lead to low template recognition accuracies, most likely due to building graph for files with a high percentage of misclassified regions.
As mentioned previously, the high v-measures reached by all strategies at a threshold of 1 is distorted
due to perfect homogeneity (all files are clustered individually).}
Surprisingly, for lower thresholds, using gold standard regions does not lead to better results.
We attribute this effect to the increased complexity of the graphs produced with sub-optimal regions: as there may be potentially more automatically detected regions than needed, the resulting graphs contain more (noisy) information and therefore show a greater absolute difference in the case of different templates.


\subsection{\change{
Scalability of template inference}}
\change{
Different region detection strategies not only have an influence on the effectiveness of \systemname's template inference step, but also affect its complexity, measurable on the run time.}
We report the execution times for the template recognition task in Table~\ref{tab:template-times}, obtained as the average run-times of our Python~3.8 scripts across three separate runs on a machine equipped with an AMD Epyc~9 7702P Xeon 3,35~GHz CPU and 512~GB of RAM.
\change{These results highlight the trade off between template inference accuracy and complexity:}
the region detection strategies that proved to be better for template inference in Figure~\ref{fig:template-scenarios} are also the ones that need significantly more time to execute, while \change{
using the region detection results of the genetic-based and TableSense approaches leads to lower running times and more imprecise results.
In fact, when incorrectly detecting regions, \systemname has a higher number of graph regions due to its partitioning steps: larger graphs need greater time for computation, but lead to more precise similarity estimates. 
The slowest runtimes on \FUSTE are obtained by \systemname in the dynamic radius scenario, because of a few files containing a large number of nodes (above 200) that lead to expensive graph similarity computations.
For this dataset, both the static radius and connected component strategies are faster, because having a fixed radius and no region partitioning lead to fewer detected regions.
Comparing \systemname across datasets, the runtimes on \DECO are lower: as this dataset is characterized by fewer templates, more file pairs with no similar region are pruned.
The same pruning strategy is less effective for the connected components strategy on \DECO, because without a clustering stage there are more spuriously similar regions across files.
}
\begin{table}[t]\small
\begin{tabular}{lrrrr}
\textbf{Region detection} & \multicolumn{4}{c}{\textbf{Template inference time} (s)} \\
 & \multicolumn{2}{c}{\DECO} & \multicolumn{2}{c}{\FUSTE} \\
\toprule
Gold Standard & \numprint{93.39}   $\pm$ & \numprint{0.26} & \numprint{78.87} $\pm$ & \numprint{0.77}\\
Dynamic Radius & \numprint{1563.51}  $\pm$ & \numprint{2.91} & \numprint{8515.46} $\pm$ & \numprint{194.55}\\
Static Radius & \numprint{343.13}  $\pm$ & \numprint{3.81} & \numprint{2749.20} $\pm$ & \numprint{13.04} \\
Connected Components & \numprint{15887.50} $\pm$ & \numprint{127.12} & \numprint{3529.21} $\pm$ & \numprint{76.67}\\
Genetic (XLS) & \numprint{102.32} $\pm$  & \numprint{0.51} & \numprint{75.12} $\pm$ & \numprint{0.96} \\
Genetic (CSV) & \numprint{114.76} $\pm$  & \numprint{1.58} &  \numprint{75.13} $\pm$ & \numprint{0.34} \\
Tablesense & \numprint{361.46} $\pm$  & \numprint{47.47} &  \numprint{51.54} $\pm$ & \numprint{9.37} \\
\bottomrule
\end{tabular}
\caption{\change{Time performance of template inference.}}
\label{tab:template-times}
\vspace{-20pt}
\end{table}
\begin{figure}[t]
    \centering
    \includegraphics[width=\linewidth]{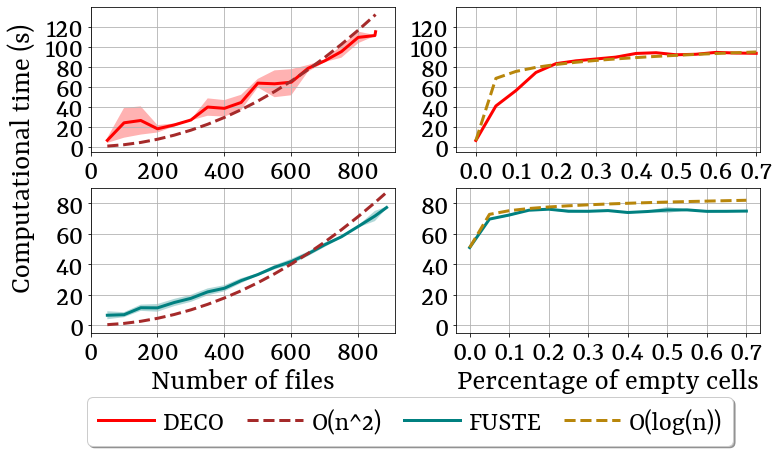}
    \caption{Effect of number of files and empty cells.}
    \label{fig:template-time}
\vspace{-2em}
\end{figure}
Figure~\ref{fig:template-time} shows the influence of the number of files and percentage of empty cells in files on the computational time of template detection using perfectly recognized regions.
For the former, we experimented by selecting random file sub-samples, with a step size of 50. 
For the latter, the sub-samples corresponded to all files with a number of empty cells up to a given percentage of the total file area, with a step size of 0.05\%.
In both cases, the file sets were sampled without repetition until full coverage of the dataset.
The plot shows that the performances with respect to the number of input files follow a quadratic behavior, as \systemname performs layout comparison for each pair of files in the input set.
In turn, increasing the percentage of empty cells leads to a logarithmic behavior.
Therefore, we conclude that the most impactful factor affecting the complexity of \systemname is the number of input files, as well as the correctness of the region detection stage:
detecting regions and multiregion file templates automatically with \systemname provides a convenient trade off between complexity and correctness.


\section{Related Work}
\label{sec:relatedwork}

While there is a substantial amount of research aimed at detecting and recognizing tables in single files of various formats, there is no research to recognize structural templates spanning different multiregion files.
The two systems WebSmatch~\cite{coletta_public_2012} and TableSense~\cite{dongLHFZ19}, like Mondrian, leverage the intuition of analyzing spreadsheets applying techniques from the computer vision domain.
The first is an internet application that uses connected component detection and machine learning for table recognition to integrate semantically related tables within a dataset.
The second uses a convolutional neural network architecture to address spreadsheet table detection based on a set of spreadsheet-specific cell features.
As demonstrated by the experimental results in Section~\ref{sec:evaluation}, supervised machine learning necessitates large quantities of training data, while the unsupervised nature of \systemname makes it fit even for smaller datasets.

Supervised learning is also used in Pytheas, by Christodoulakis et al.~\cite{pytheas20}.
This system employs a rule-based algorithm to discover tables in .csv files. 
Due to the nature of .csv files, tabular structures are expected to appear concatenated in one dimension, i.e., as subsequent lines (or columns).
In contrast, \systemname can detect region layouts with an extra degree of freedom, recognizing both horizontal and vertical alignments (cf.~Figure~\ref{fig:rendering}).

Encoding tabular layouts as graphs is at the core of the approach presented by Koci et al.~\cite{koci2019genetic}, which tackles table recognition in spreadsheet files with a combination of supervised machine learning and genetic-based algorithms.
While \systemname uses complete graphs that encode all regions in a file with their pairwise distances, the genetic-based approach focuses on tabular regions and therefore misses information about the general file layout.

Existing spreadsheet systems that build on region boundary extraction can integrate well with \systemname and make use of its layout template recognition.
For example, to perform information extraction, the work of Chen et al.~\cite{chen2017Property}, given table boundaries, leverages active learning to detect interesting ``spreadsheet properties'', such as aggregation rows or hierarchies.
Detecting spreadsheet templates can lead to a significant reduction of the number of files for which user feedback is required.
Spreadsheet data management systems, like Senbazuru~\cite{chenSenbazuruPrototypeSpreadsheet2013}, can be empowered with \systemname, e.g., using layout templates as database indices, or enriching query results with template information.
\section{Conclusions}
\label{sec:final}

In this work,
we formalized a framework for describing \multifile layout templates and identified three main challenges: detecting independent region boundaries in a single spreadsheet; matching similar regions on a structural level; finding a suitable similarity for file layouts.
We presented the \systemname approach, which combines automated region detection with an algorithm to identify similar file layouts.
Experiments show that our approach works well in detecting the boundaries of different regions in a \multifile spreadsheet and
in identifying layout templates.
Further research will focus on improving the accuracy of boundary detection and increasing the quality of the detected file layouts.
We plan to include more information in the structure similarity computation, e.g., a finer-grained classification for the content of spreadsheet cells, to better identify structural patterns and correlations within templates.
\balance





\begin{acks}
This research was funded by the HPI research school
on Data Science and Engineering.
\end{acks}


\bibliographystyle{ACM-Reference-Format}
\bibliography{references}


\providecommand{\noopsort}[1]{}
\begin{thebibliography}{26}


\ifx \showCODEN    \undefined \def \showCODEN     #1{\unskip}     \fi
\ifx \showDOI      \undefined \def \showDOI       #1{#1}\fi
\ifx \showISBNx    \undefined \def \showISBNx     #1{\unskip}     \fi
\ifx \showISBNxiii \undefined \def \showISBNxiii  #1{\unskip}     \fi
\ifx \showISSN     \undefined \def \showISSN      #1{\unskip}     \fi
\ifx \showLCCN     \undefined \def \showLCCN      #1{\unskip}     \fi
\ifx \shownote     \undefined \def \shownote      #1{#1}          \fi
\ifx \showarticletitle \undefined \def \showarticletitle #1{#1}   \fi
\ifx \showURL      \undefined \def \showURL       {\relax}        \fi
\providecommand\bibfield[2]{#2}
\providecommand\bibinfo[2]{#2}
\providecommand\natexlab[1]{#1}
\providecommand\showeprint[2][]{arXiv:#2}

\bibitem[\protect\citeauthoryear{Bajuelos, Tom{\'a}s, and Marques}{Bajuelos
  et~al\mbox{.}}{2004}]%
        {bajuelos2004partitioning}
\bibfield{author}{\bibinfo{person}{Ant{\'o}nio~Leslie Bajuelos},
  \bibinfo{person}{Ana~Paula Tom{\'a}s}, {and} \bibinfo{person}{F{\'a}bio
  Marques}.} \bibinfo{year}{2004}\natexlab{}.
\newblock \showarticletitle{Partitioning orthogonal polygons by extension of
  all edges incident to reflex vertices: lower and upper bounds on the number
  of pieces}. In \bibinfo{booktitle}{\emph{International Conference on
  Computational Science and Its Applications (ICCSA)}}.
  \bibinfo{pages}{127--136}.
\newblock


\bibitem[\protect\citeauthoryear{Chen, Cafarella, Chen, Prevo, and Zhuang}{Chen
  et~al\mbox{.}}{2013}]%
        {chenSenbazuruPrototypeSpreadsheet2013}
\bibfield{author}{\bibinfo{person}{Zhe Chen}, \bibinfo{person}{Michael
  Cafarella}, \bibinfo{person}{Jun Chen}, \bibinfo{person}{Daniel Prevo}, {and}
  \bibinfo{person}{Junfeng Zhuang}.} \bibinfo{year}{2013}\natexlab{}.
\newblock \showarticletitle{Senbazuru: A Prototype Spreadsheet Database
  Management System}.
\newblock \bibinfo{journal}{\emph{PVLDB}} \bibinfo{volume}{6},
  \bibinfo{number}{12} (\bibinfo{year}{2013}), \bibinfo{pages}{1202--1205}.
\newblock


\bibitem[\protect\citeauthoryear{Chen and Cafarella}{Chen and
  Cafarella}{2013}]%
        {chenCRF}
\bibfield{author}{\bibinfo{person}{Zhe Chen} {and} \bibinfo{person}{Michael~J.
  Cafarella}.} \bibinfo{year}{2013}\natexlab{}.
\newblock \showarticletitle{Automatic web spreadsheet data extraction}. In
  \bibinfo{booktitle}{\emph{International Workshop on Semantic Search over the
  Web (SSW)}}. \bibinfo{pages}{1:1--1:8}.
\newblock


\bibitem[\protect\citeauthoryear{Chen, Dadiomov, Wesley, Xiao, Cory, Cafarella,
  and Mackinlay}{Chen et~al\mbox{.}}{2017}]%
        {chen2017Property}
\bibfield{author}{\bibinfo{person}{Zhe Chen}, \bibinfo{person}{Sasha Dadiomov},
  \bibinfo{person}{Richard Wesley}, \bibinfo{person}{Gang Xiao},
  \bibinfo{person}{Daniel Cory}, \bibinfo{person}{Michael~J. Cafarella}, {and}
  \bibinfo{person}{Jock~D. Mackinlay}.} \bibinfo{year}{2017}\natexlab{}.
\newblock \showarticletitle{Spreadsheet Property Detection With Rule-assisted
  Active Learning}. In \bibinfo{booktitle}{\emph{Proceedings of the
  International Conference on Information and Knowledge Management (CIKM)}}.
  \bibinfo{pages}{999--1008}.
\newblock


\bibitem[\protect\citeauthoryear{Chiticariu, Li, Raghavan, and
  Reiss}{Chiticariu et~al\mbox{.}}{2010}]%
        {chiticariuEnterpriseInformationExtraction2010}
\bibfield{author}{\bibinfo{person}{Laura Chiticariu}, \bibinfo{person}{Yunyao
  Li}, \bibinfo{person}{Sriram Raghavan}, {and} \bibinfo{person}{Frederick~R.
  Reiss}.} \bibinfo{year}{2010}\natexlab{}.
\newblock \showarticletitle{Enterprise Information Extraction: Recent
  Developments and Open Challenges}. In \bibinfo{booktitle}{\emph{Proceedings
  of the International Conference on Management of Data (SIGMOD)}}.
  \bibinfo{pages}{1257--1258}.
\newblock


\bibitem[\protect\citeauthoryear{Christodoulakis, Munson, Gabel, Brown, and
  Miller}{Christodoulakis et~al\mbox{.}}{2020}]%
        {pytheas20}
\bibfield{author}{\bibinfo{person}{Christina Christodoulakis},
  \bibinfo{person}{Eric Munson}, \bibinfo{person}{Moshe Gabel},
  \bibinfo{person}{Angela~Demke Brown}, {and} \bibinfo{person}{Ren{\'{e}}e~J.
  Miller}.} \bibinfo{year}{2020}\natexlab{}.
\newblock \showarticletitle{Pytheas: Pattern-based Table Discovery in {CSV}
  Files}.
\newblock \bibinfo{journal}{\emph{PVLDB}} \bibinfo{volume}{13},
  \bibinfo{number}{11} (\bibinfo{year}{2020}), \bibinfo{pages}{2075--2089}.
\newblock


\bibitem[\protect\citeauthoryear{Coletta, Castanier, Valduriez, Frisch, Ngo,
  and Bellahsene}{Coletta et~al\mbox{.}}{2012}]%
        {coletta_public_2012}
\bibfield{author}{\bibinfo{person}{Remi Coletta}, \bibinfo{person}{Emmanuel
  Castanier}, \bibinfo{person}{Patrick Valduriez}, \bibinfo{person}{Christian
  Frisch}, \bibinfo{person}{DuyHoa Ngo}, {and} \bibinfo{person}{Zohra
  Bellahsene}.} \bibinfo{year}{2012}\natexlab{}.
\newblock \showarticletitle{Public data integration with {WebSmatch}}. In
  \bibinfo{booktitle}{\emph{Proceedings of the {International} {Workshop} on
  {Open} {Data} ({WOD})}}. \bibinfo{pages}{5--12}.
\newblock


\bibitem[\protect\citeauthoryear{Ester, Kriegel, Sander, and Xu}{Ester
  et~al\mbox{.}}{1996}]%
        {dbscan1996}
\bibfield{author}{\bibinfo{person}{Martin Ester}, \bibinfo{person}{Hans-Peter
  Kriegel}, \bibinfo{person}{J{\"o}rg Sander}, {and} \bibinfo{person}{Xiaowei
  Xu}.} \bibinfo{year}{1996}\natexlab{}.
\newblock \showarticletitle{A density-based algorithm for discovering clusters
  in large spatial databases with noise}. In
  \bibinfo{booktitle}{\emph{Proceedings of the International Conference on
  Knowledge Discovery and Data Mining (SIGKDD)}}. \bibinfo{pages}{5--12}.
\newblock


\bibitem[\protect\citeauthoryear{Fisher and Rothermel}{Fisher and
  Rothermel}{2005}]%
        {euses2005}
\bibfield{author}{\bibinfo{person}{Marc Fisher} {and} \bibinfo{person}{Gregg
  Rothermel}.} \bibinfo{year}{2005}\natexlab{}.
\newblock \showarticletitle{The {EUSES} spreadsheet corpus: a shared resource
  for supporting experimentation with spreadsheet dependability mechanisms}.
\newblock \bibinfo{journal}{\emph{{ACM} {SIGSOFT} Software Engineering Notes}}
  \bibinfo{volume}{30}, \bibinfo{number}{4} (\bibinfo{year}{2005}),
  \bibinfo{pages}{1--5}.
\newblock


\bibitem[\protect\citeauthoryear{Gilani, Qasim, Malik, and Shafait}{Gilani
  et~al\mbox{.}}{2017}]%
        {gilani2017table}
\bibfield{author}{\bibinfo{person}{Azka Gilani}, \bibinfo{person}{Shah~Rukh
  Qasim}, \bibinfo{person}{Imran Malik}, {and} \bibinfo{person}{Faisal
  Shafait}.} \bibinfo{year}{2017}\natexlab{}.
\newblock \showarticletitle{Table detection using deep learning}. In
  \bibinfo{booktitle}{\emph{Proceedings of the IAPR International Conference on
  Document Analysis and Recognition (ICDAR)}}. \bibinfo{publisher}{{IEEE}},
  \bibinfo{address}{Washington, DC, USA}, \bibinfo{pages}{771--776}.
\newblock


\bibitem[\protect\citeauthoryear{Haoyu, Shijie, Shi, Zhouyu, and Dongmei}{Haoyu
  et~al\mbox{.}}{2019}]%
        {dongLHFZ19}
\bibfield{author}{\bibinfo{person}{Dong Haoyu}, \bibinfo{person}{Liu Shijie},
  \bibinfo{person}{Han Shi}, \bibinfo{person}{Fu Zhouyu}, {and}
  \bibinfo{person}{Zhang Dongmei}.} \bibinfo{year}{2019}\natexlab{}.
\newblock \showarticletitle{TableSense: Spreadsheet Table Detection with
  Convolutional Neural Networks}. In \bibinfo{booktitle}{\emph{Proceedings of
  the AAAI Conference on Artificial Intelligence (AAAI)}}.
  \bibinfo{pages}{69--76}.
\newblock


\bibitem[\protect\citeauthoryear{He, Gkioxari, Doll{\'{a}}r, and Girshick}{He
  et~al\mbox{.}}{2017}]%
        {heGDG17MaskRCNN}
\bibfield{author}{\bibinfo{person}{Kaiming He}, \bibinfo{person}{Georgia
  Gkioxari}, \bibinfo{person}{Piotr Doll{\'{a}}r}, {and}
  \bibinfo{person}{Ross~B. Girshick}.} \bibinfo{year}{2017}\natexlab{}.
\newblock \showarticletitle{Mask {R-CNN}}. In
  \bibinfo{booktitle}{\emph{Proceedings of the {IEEE} International Conference
  on Computer Vision (ICCV)}}. \bibinfo{pages}{2980--2988}.
\newblock


\bibitem[\protect\citeauthoryear{Hermans and {Murphy-Hill}}{Hermans and
  {Murphy-Hill}}{2015}]%
        {enron2015}
\bibfield{author}{\bibinfo{person}{Felienne Hermans} {and}
  \bibinfo{person}{Emerson {Murphy-Hill}}.} \bibinfo{year}{2015}\natexlab{}.
\newblock \showarticletitle{Enron's {{Spreadsheets}} and {{Related Emails}}:
  {{A Dataset}} and {{Analysis}}}. In \bibinfo{booktitle}{\emph{Proceedings of
  the International Conference on Software Engineering (ICSE)}}.
  \bibinfo{pages}{7--16}.
\newblock


\bibitem[\protect\citeauthoryear{Koci, Thiele, Lehner, and Romero}{Koci
  et~al\mbox{.}}{2018}]%
        {koci2018graph}
\bibfield{author}{\bibinfo{person}{Elvis Koci}, \bibinfo{person}{Maik Thiele},
  \bibinfo{person}{Wolfgang Lehner}, {and} \bibinfo{person}{Oscar Romero}.}
  \bibinfo{year}{2018}\natexlab{}.
\newblock \showarticletitle{Table recognition in spreadsheets via a graph
  representation}. In \bibinfo{booktitle}{\emph{Proceedings of the IAPR
  International Workshop on Document Analysis Systems (DAS)}}.
  \bibinfo{pages}{139--144}.
\newblock


\bibitem[\protect\citeauthoryear{Koci, Thiele, Rehak, Romero, and Lehner}{Koci
  et~al\mbox{.}}{2019a}]%
        {koci2019deco}
\bibfield{author}{\bibinfo{person}{Elvis Koci}, \bibinfo{person}{Maik Thiele},
  \bibinfo{person}{Josephine Rehak}, \bibinfo{person}{Oscar Romero}, {and}
  \bibinfo{person}{Wolfgang Lehner}.} \bibinfo{year}{2019}\natexlab{a}.
\newblock \showarticletitle{{DECO}: A Dataset of Annotated Spreadsheets for
  Layout and Table Recognition}. In \bibinfo{booktitle}{\emph{Proceedings of
  the IAPR International Conference on Document Analysis and Recognition
  (ICDAR)}}. \bibinfo{pages}{1280--1285}.
\newblock


\bibitem[\protect\citeauthoryear{Koci, Thiele, Romero, and Lehner}{Koci
  et~al\mbox{.}}{2016}]%
        {koci2016machine}
\bibfield{author}{\bibinfo{person}{Elvis Koci}, \bibinfo{person}{Maik Thiele},
  \bibinfo{person}{Oscar Romero}, {and} \bibinfo{person}{Wolfgang Lehner}.}
  \bibinfo{year}{2016}\natexlab{}.
\newblock \showarticletitle{A Machine Learning Approach for Layout Inference in
  Spreadsheets}. In \bibinfo{booktitle}{\emph{Proceedings of the
  {International} {Joint} {Conference} on {Knowledge} {Discovery}, {Knowledge}
  {Engineering} and {Knowledge} {Management} (IC3K)}}. \bibinfo{pages}{77--88}.
\newblock


\bibitem[\protect\citeauthoryear{Koci, Thiele, Romero, and Lehner}{Koci
  et~al\mbox{.}}{2019b}]%
        {koci2019genetic}
\bibfield{author}{\bibinfo{person}{Elvis Koci}, \bibinfo{person}{Maik Thiele},
  \bibinfo{person}{Oscar Romero}, {and} \bibinfo{person}{Wolfgang Lehner}.}
  \bibinfo{year}{2019}\natexlab{b}.
\newblock \showarticletitle{A Genetic-based Search for Adaptive Table
  Recognition in Spreadsheets}. In \bibinfo{booktitle}{\emph{Proceedings of the
  IAPR International Conference on Document Analysis and Recognition (ICDAR)}}.
  \bibinfo{pages}{1274--1279}.
\newblock


\bibitem[\protect\citeauthoryear{Koutras, Siachamis, Ionescu, Psarakis, Brons,
  Fragkoulis, Lofi, Bonifati, and Katsifodimos}{Koutras et~al\mbox{.}}{2021}]%
        {koutrasSIPBFLBK21}
\bibfield{author}{\bibinfo{person}{Christos Koutras}, \bibinfo{person}{George
  Siachamis}, \bibinfo{person}{Andra Ionescu}, \bibinfo{person}{Kyriakos
  Psarakis}, \bibinfo{person}{Jerry Brons}, \bibinfo{person}{Marios
  Fragkoulis}, \bibinfo{person}{Christoph Lofi}, \bibinfo{person}{Angela
  Bonifati}, {and} \bibinfo{person}{Asterios Katsifodimos}.}
  \bibinfo{year}{2021}\natexlab{}.
\newblock \showarticletitle{Valentine: Evaluating Matching Techniques for
  Dataset Discovery}. In \bibinfo{booktitle}{\emph{Proceedings of the
  International Conference on Data Engineering (ICDE)}}.
  \bibinfo{pages}{468--479}.
\newblock


\bibitem[\protect\citeauthoryear{Melnik, Garcia-Molina, and Rahm}{Melnik
  et~al\mbox{.}}{2001}]%
        {melnikGR02extended}
\bibfield{author}{\bibinfo{person}{Sergey Melnik}, \bibinfo{person}{Hector
  Garcia-Molina}, {and} \bibinfo{person}{Erhard Rahm}.}
  \bibinfo{year}{2001}\natexlab{}.
\newblock \bibinfo{booktitle}{\emph{Similarity Flooding: A Versatile Graph
  Matching Algorithm (Extended Technical Report)}}.
\newblock \bibinfo{type}{Technical Report} 2001-25.
  \bibinfo{institution}{Stanford InfoLab}.
\newblock
\urldef\tempurl%
\url{http://ilpubs.stanford.edu:8090/497/}
\showURL{%
\tempurl}


\bibitem[\protect\citeauthoryear{Melnik, Garcia{-}Molina, and Rahm}{Melnik
  et~al\mbox{.}}{2002}]%
        {melnikGR02}
\bibfield{author}{\bibinfo{person}{Sergey Melnik}, \bibinfo{person}{Hector
  Garcia{-}Molina}, {and} \bibinfo{person}{Erhard Rahm}.}
  \bibinfo{year}{2002}\natexlab{}.
\newblock \showarticletitle{Similarity Flooding: {A} Versatile Graph Matching
  Algorithm and Its Application to Schema Matching}. In
  \bibinfo{booktitle}{\emph{Proceedings of the International Conference on Data
  Engineering (ICDE)}}. \bibinfo{pages}{117--128}.
\newblock


\bibitem[\protect\citeauthoryear{Mitl{\"o}hner, Neumaier, Umbrich, and
  Polleres}{Mitl{\"o}hner et~al\mbox{.}}{2016}]%
        {openCSV2016}
\bibfield{author}{\bibinfo{person}{Johann Mitl{\"o}hner},
  \bibinfo{person}{Sebastian Neumaier}, \bibinfo{person}{J{\"u}rgen Umbrich},
  {and} \bibinfo{person}{Axel Polleres}.} \bibinfo{year}{2016}\natexlab{}.
\newblock \showarticletitle{Characteristics of open data CSV files}. In
  \bibinfo{booktitle}{\emph{Proceedings of the Image Analysis and Processing
  Conference ({ICIAP})}}. \bibinfo{pages}{72--79}.
\newblock


\bibitem[\protect\citeauthoryear{Nargesian, Zhu, Miller, Pu, and
  Arocena}{Nargesian et~al\mbox{.}}{2019}]%
        {nargesian2019datalake}
\bibfield{author}{\bibinfo{person}{Fatemeh Nargesian}, \bibinfo{person}{Erkang
  Zhu}, \bibinfo{person}{Ren{\'{e}}e~J. Miller}, \bibinfo{person}{Ken~Q. Pu},
  {and} \bibinfo{person}{Patricia~C. Arocena}.}
  \bibinfo{year}{2019}\natexlab{}.
\newblock \showarticletitle{Data Lake Management: Challenges and
  Opportunities}.
\newblock \bibinfo{journal}{\emph{PVLDB}} (\bibinfo{year}{2019}).
\newblock


\bibitem[\protect\citeauthoryear{Rosenberg and Hirschberg}{Rosenberg and
  Hirschberg}{2007}]%
        {vmeasure07}
\bibfield{author}{\bibinfo{person}{Andrew Rosenberg} {and}
  \bibinfo{person}{Julia Hirschberg}.} \bibinfo{year}{2007}\natexlab{}.
\newblock \showarticletitle{V-Measure: {A} Conditional Entropy-Based External
  Cluster Evaluation Measure}. In \bibinfo{booktitle}{\emph{Proceedings of the
  Joint Conference on Empirical Methods in Natural Language Processing and
  Computational Natural Language Learning (EMNLP-CoNLL)}}.
  \bibinfo{pages}{410--420}.
\newblock


\bibitem[\protect\citeauthoryear{Titus, Kevin, Justin, John, and
  Emerson~R.}{Titus et~al\mbox{.}}{2015}]%
        {fuse2015}
\bibfield{author}{\bibinfo{person}{Barik Titus}, \bibinfo{person}{Lubick
  Kevin}, \bibinfo{person}{Smith Justin}, \bibinfo{person}{Slankas John}, {and}
  \bibinfo{person}{Murphy{-}Hill Emerson~R.}} \bibinfo{year}{2015}\natexlab{}.
\newblock \showarticletitle{Fuse: {A} Reproducible, Extendable, Internet-Scale
  Corpus of Spreadsheets}. In \bibinfo{booktitle}{\emph{{IEEE/ACM} Working
  Conference on Mining Software Repositories, {MSR}}}.
  \bibinfo{pages}{486--489}.
\newblock


\bibitem[\protect\citeauthoryear{Zhang and Ives}{Zhang and Ives}{2020}]%
        {zhangI20}
\bibfield{author}{\bibinfo{person}{Yi Zhang} {and} \bibinfo{person}{Zachary~G.
  Ives}.} \bibinfo{year}{2020}\natexlab{}.
\newblock \showarticletitle{Finding Related Tables in Data Lakes for
  Interactive Data Science}. In \bibinfo{booktitle}{\emph{Proceedings of the
  International Conference on Management of Data (SIGMOD)}}.
  \bibinfo{pages}{1951--1966}.
\newblock


\bibitem[\protect\citeauthoryear{Zhu, Deng, Nargesian, and Miller}{Zhu
  et~al\mbox{.}}{2019}]%
        {zhuDNM19}
\bibfield{author}{\bibinfo{person}{Erkang Zhu}, \bibinfo{person}{Dong Deng},
  \bibinfo{person}{Fatemeh Nargesian}, {and} \bibinfo{person}{Ren{\'{e}}e~J.
  Miller}.} \bibinfo{year}{2019}\natexlab{}.
\newblock \showarticletitle{{JOSIE:} Overlap Set Similarity Search for Finding
  Joinable Tables in Data Lakes}. In \bibinfo{booktitle}{\emph{Proceedings of
  the International Conference on Management of Data (SIGMOD)}}.
\newblock


\end{thebibliography}
\balance
\end{document}